\documentclass[journal,onecolumn]{IEEEtran}
\pdfoutput=1 
\usepackage{cite}
\usepackage{setspace}
\usepackage{amsmath,amssymb,amsfonts}
\usepackage{graphicx}
\usepackage{textcomp}
\usepackage{xcolor}
\usepackage[]{algorithm}%
\usepackage{enumitem}
\usepackage{algcompatible}
\usepackage{tabularray}
\usepackage{threeparttable}
\usepackage[hidelinks]{hyperref}
\newcommand\copyrighttext{%
	\footnotesize \textcopyright 2023 IEEE. 
	Personal use of this material is permitted. Permission from IEEE must be obtained for all other uses, in any current or future media, including reprinting/republishing this material for advertising or promotional purposes, creating new collective works, for resale or redistribution to servers or lists, or reuse of any copyrighted component of this work in other works. 
	DOI: \href{https://ieeexplore.ieee.org/document/10210076}{10.1109/TSIPN.2023.3302658}
}
\makeatletter
\def\ps@IEEEtitlepagestyle{
	\def\@oddfoot{\mycopyrightnotice}
	\def\@evenfoot{}
}
\def\mycopyrightnotice{
	{\footnotesize
		\begin{minipage}{\textwidth-2\fboxsep}%
			\centering%
			\noindent\fbox{\parbox{\linewidth}{\copyrighttext}}
		\end{minipage}
	}
}
\usepackage{pifont}

\algblockdefx{FORALLP}{ENDFAP}[1]%
{\textbf{for all }#1 \textbf{do in parallel}}%
{\textbf{end for}}
\algloopdefx{RETURN}[1][]{\textbf{return} #1}

\algblockdefx{FORALLN}{ENDFAN}[1]%
{\textbf{for all }#1 \textbf{do in each node}}%
{\textbf{end for}}
\algloopdefx{RETURN}[1][]{\textbf{return} #1}

\usepackage{nccmath}
\usepackage{lipsum}
\usepackage[nolist]{acronym}
\usepackage{gensymb}
\floatname{algorithm}{Algorithm}

\newcommand{\myvspace}{\vspace{0pt}}

\newtheorem{remark}{Remark}
  \usepackage{pgfplots}
  \usepackage{pgfplotstable}
  \pgfplotsset{compat=newest}
  \usetikzlibrary{plotmarks}
  \usetikzlibrary{arrows.meta}
  \usepgfplotslibrary{patchplots}
  \usepackage{grffile}
  \usetikzlibrary{spy,backgrounds}
  \usetikzlibrary{calc}
  
\usepackage{tikz}

\definecolor{lime}{HTML}{A6CE39}
\DeclareRobustCommand{\orcidicon}{
	\begin{tikzpicture}
		\draw[lime, fill=lime] (0,0) 
		circle [radius=0.16] 
		node[white] {{\fontfamily{qag}\selectfont \tiny ID}};
		\draw[white, fill=white] (-0.0625,0.095) 
		circle [radius=0.007];
	\end{tikzpicture}
	\hspace{-2mm}
}

\foreach \x in {A, ..., Z}{\expandafter\xdef\csname orcid\x\endcsname{\noexpand\href{https://orcid.org/\csname orcidauthor\x\endcsname}
		{\noexpand\orcidicon}}
}

\newcommand{\diag}[1]{\operatorname{diag}\left( #1 \right)}

\newtheorem{theorem}{Theorem}

\begin{document}
\pagestyle{empty}

\title{Decentralized Eigendecomposition for Online Learning over Graphs with Applications}

\author{{Yufan Fan, Minh Trinh-Hoang, Cemil Emre Ardic and Marius Pesavento}
\thanks{This paper was presented in part at the 29th European Signal Processing Conference, Dublin, Ireland, August 23-27, 2021, and in part at the 12th IEEE Sensor Array and Multichannel Signal Processing Workshop, Trondheim, Norway, June 20-23, 2022. (\textit{Corresponding author: Yufan Fan.})}
\thanks{Y. Fan, M. Trinh-Hoang, C. Ardic and M. Pesavento are with the Communication Systems Group, Technische Universit\"at Darmstadt, Darmstadt 64283, Germany (e-mail: yufan.fan@nt.tu-darmstadt.de; thminh@nt.tu-darmstadt.de; cemil.emre.ardic@izfp.fraunhofer.de; pesavento@nt.tu-darmstadt.de).}
\thanks{This work was financially supported in part by the Federal Ministry of Education and Research of Germany in the project ``Open6GHub" (grant no 16KISK014), and in part by the German Research Foundation in the project ``PRIDE" (grant no PE2080/2-1).}
\thanks{\copyrighttext}
}

\maketitle

\thispagestyle{empty}

\begin{acronym}[PR-DML]
	 \acro{ULA}{Uniform Linear Array}
	 \acro{DoA}{Direction-of-Arrival}
	 \acro{FoV}{Field of View}
	 \acro{MUSIC}{Multiple Signal Classification}
	 \acro{ESPRIT}{Estimation of Signal Parameters via Rotational Invariance Technique}
	 \acro{DML}{Deterministic Maximum Likelihood}
	 \acro{SML}{Stochastic Maximum Likelihood}
	 \acro{WSF}{Weighted Subspace Fitting}
	 \acro{CF}{Covariance Fitting}
	 \acro{FCF}{Full Covariance Fitting}
	 \acro{PR}{Partial Relaxation}
	 \acro{PR-DML}{Partially Relaxed Deterministic Maximum Likelihood}
	 \acro{PR-CF}{Partially Relaxed Covariance Fitting}
	 \acro{PR-WSF}{Partially Relaxed Weighted Subspace Fitting}
	 \acro{MDL}{Minimum Description Length}
	 \acro{SNR}{Signal-to-Noise Ratio} 
	 \acro{LS}{Least Square}
	 \acro{ML}{Maximum Likelihood}
	 \acro{LR}{Likelihood Ratio}
	 \acro{FWE}{Familywise Error-Rate}
	 \acro{FDR}{False Discovery Rate}
	 \acro{RMT}{Random Matrix Theory}
	 \acro{CRB}{Cramer-Rao Bound}
	 \acro{RMSE}{Root-Mean-Squared-Error}
	 \acro{DCT}{Dominated Convergence Theorem}
	 \acro{pdf}{probability density function}
	 \acro{cdf}{cumulative distribution function}
	 \acro{MSE}{Mean Square Error}
	 \acro{CLT}{Central Limit Theorem}
	 \acro{FD}{Fourier Domain}
	 \acro{DFT}{Discrete Fourier Transform}
\end{acronym}

\begin{abstract}
In this article, the problem of decentralized eigenvalue decomposition of a general symmetric matrix that is important, e.g., in Principal Component Analysis, is studied, and a decentralized online learning algorithm is proposed. Instead of collecting all information in a fusion center, the proposed algorithm involves only local interactions among adjacent agents. It benefits from the representation of the matrix as a sum of rank-one components which makes the algorithm attractive for online eigenvalue and eigenvector tracking applications. We examine the performance of the proposed algorithm in two types of important application examples: First, we consider the online eigendecomposition of a sample covariance matrix over the network, with application in decentralized Direction-of-Arrival (DoA) estimation and DoA tracking applications. Then, we investigate the online computation of the spectra of the graph Laplacian that is important in, e.g., Graph Fourier Analysis and graph dependent filter design. We apply our proposed algorithm to track the spectra of the graph Laplacian in static and dynamic networks. Simulation results reveal that the proposed algorithm outperforms existing decentralized algorithms both in terms of estimation accuracy as well as communication cost.
\end{abstract}

\begin{IEEEkeywords}
	Graph signal processing, decentralized online algorithm, decentralized eigendecomposition, rank-one modification problem, rational function approximation, graph filters, DoA estimation.
\end{IEEEkeywords}

\section{Motivation and Introduction}\label{sec:sysMod}

%
%
%
%
%
%
%
%
%
%
%

The eigenvalue decomposition is fundamental in various application areas such as signal processing, data mining, and machine learning\cite{a102,a103,a104}. Particularly when the dimension of the data is large, dimensionality reduction techniques, such as the Principal Component Analysis (PCA), are required to obtain lower-dimensional representations of the data, e.g., by means of eigendecomposition \cite{a95,a96,a97}. Moreover, scalable solutions, such as distributed algorithms, are of high interest in big data and machine learning applications, and when the data, e.g., local sensor measurements \cite{a146}, anonymous surveys \cite{a147}, private statistics \cite{a148}, is massively distributed over a network of agents. Based on the concept of in-network processing, instead of collecting all information in a fusion center, agents perform local processing and collaborate by exchanging information only locally with their neighbors, and the significant bandwidth requirement at the fusion center is avoided. Such decentralized systems usually benefit from the robustness to, e.g., agent failure, which could lead to a complete breakdown of the system, and from the scalability in the sense that via collaboration a single agent is not limited to its own storage and computation resources \cite{a149}.

The decentralized PCA algorithms can be categorized into two classes based on how the measurement matrix is partitioned over the network, i.e., the sample-wise partitioning and the feature-wise partitioning \cite{a140,a139}. While in the sample-wise partitioning, e.g., in \cite{a142,a143,a144,a145}, each agent has access to a different subset of samples of the data set that contains the entire set of features, in the feature-wise partitioning, each node has access to all observations of a single feature (or an exclusive subset of features). This arises naturally in, for example, distributed sensor deployments and distributed antenna arrays, which is the focus of our work.

Different decentralized PCA algorithms with the feature-wise partitioning setup have been proposed in the literature by applying consensus gossiping strategies \cite{a98,a99}. Based on the Average Consensus (AC) algorithm the decentralized Power Method (d-PM) is presented in \cite{a91}. The distributed Normalized Oja's (d-Oja) method for distributed subspace estimation is introduced in \cite{a100}, where the Oja's rule is performed using only local interactions. The Distributed Adaptive Covariance Matrix Eigenvector Estimation (DACMEE) algorithm is proposed in \cite{a101} for distributed eigenvector computation of a sample covariance matrix, whose applicability is, however, limited to fully connected or tree network topologies. Recently, the authors in \cite{a139} proposed a distributed PCA algorithm that combines a variant of the PM, i.e., the Orthogonal Iteration (OI) \cite{a116} method, and the AC algorithm to find the principal eigenspace of the covariance matrix simultaneously.

The distributed eigenvalue decomposition is valuable in various applications, such as the distributed Estimation of Signal Parameters via Rotational Invariance Techniques (d-ESPRIT) algorithm \cite{a119,a92} for decentralized Direction-of-Arrival (DoA) estimation. Combining the AC algorithm and the non-Hermitian generalized eigendecomposition, an online adaptive algorithm is proposed in \cite{a94} to perform decentralized cooperative DoA tracking. Moreover, distributed DoA tracking is carried out in \cite{a127}, which is based on a distributed implementation of the Projection Approximation Subspace Tracking (PAST) algorithm proposed in \cite{a126}.

Besides the application example in decentralized DoA estimation, the distributed eigenvalue decomposition enables the distributed inference of networks, which is important for a variety of applications in graph signal processing (GSP). More specifically, the knowledge of the eigenvalues of the shift operator, also known as the graph frequencies or the graph spectrum, plays an important role in various network inference tasks. For example, the finite-time AC (ftAC) algorithms proposed in \cite{a106,a109,a111} require the knowledge of the eigenvalues of the graph Laplacian to achieve the average consensus in finite time. Furthermore, the knowledge of the graph spectrum is required to design filters of reduced filter length in graph dependent filter designs \cite{a120,a121,a122,a123}.

In contrast to the d-PM which we take as a benchmark, in this work we propose a decentralized implementation of an online and adaptive eigendecomposition algorithm that does not rely on the power iteration but is based on the eigendecomposition of a rank-one modified diagonal matrix\cite{a10,a125}. Moreover, our proposed algorithm does not suffer from divergence as encountered by other non power iteration based algorithms, e.g., the MAximum Likelihood Adaptive Subspace Estimation (MALASE) method \cite{a128}, the OPErator Restriction Algorithm (OPERA) \cite{a129} and the Given's rotation based URV updating method \cite{a130}, even though they are centralized algorithms \cite{a131}. 

In our proposed algorithm the data available at each agent is diffused through the network using parallel consensus protocols with local interactions between agents. At termination, each agent has the knowledge of all eigenvalues and one row (or multiple rows) of the eigenvector matrix corresponding to the index of the agent. The benefit of our distributed scheme with respect to the popular d-PM is that all eigenvalues and eigenvectors are computed in parallel and that the algorithm is particularly suitable for online tracking applications where rank-one updates are natural. We evaluate the performance of our decentralized eigendecomposition algorithm in two types of prominent application examples: (A) the decentralized eigendecomposition of an evolving sample covariance matrix and (B) the decentralized online computation of the graph eigenvectors and eigenvalues in a dynamically evolving graphical network. To summarize, our contributions are as follows:
\begin{itemize}[leftmargin=19pt]
	\item We address the decentralized eigenvalue decomposition as the problem of computing the eigenvalues of a rank-one modification of a general symmetric matrix. The eigenvalues can be efficiently updated by the local rational function approximation approach \cite{a90}, which is especially suitable for decentralized online eigenvalue estimation and tracking applications.
	\item Agnostic to a specific consensus protocol, we combine the rational function approximation and any consensus protocols to propose a distributed implementation that not only consumes less total communication cost but also achieves better estimation accuracy than the state-of-the-art d-PM \cite{a91} and d-NOja method \cite{a100} for distributed eigendecomposition.
	\item We examine the application scenario where our distributed algorithm is used to perform the eigendecomposition of an evolving sample covariance matrix. This is further developed for DoA estimation and tracking.
	\item We apply our distributed algorithm to the spectrum computation and tracking of the graph Laplacian of dynamic networks with evolving topologies. This is further adopted to customize and speed up the upper level decentralized algorithms, such as the finite-time Average Consensus protocol and the graph based filter design.
\end{itemize}

The article is organized as follows. In Section \ref{sec:disAlg} we propose a distributed implementation of the online eigenvalue decomposition algorithm based on decentralized averaging protocols and the rational function approximation approach. Four different decentralized averaging approaches are briefly revised in Section \ref{sec:protocols}. In Section \ref{sec:simulation}, different application scenarios are studied, i.e., the eigenvalue decomposition of the sample covariance matrix, the decentralized DoA estimation and the online DoA tracking, the spectrum computation in dynamic graphs, the spectrum computation in dynamic graphs with rank-two updates, and the eigenvalue decomposition of the sample covariance matrix with a stabilizing adapted graph Laplacian. We then conclude our article in Section \ref{sec: conclusion}.

\textit{Notation:} The regular letter $a$ denotes a scalar, the boldface lowercase letter $\mathbf{a}$ denotes a column vector, and the boldface uppercase letter $\mathbf{A}$ denotes a matrix. The calligraphic letter $\mathcal{A}$ denotes a set. The symbols $\mathbb{R}$ and $\mathbb{C}$ represent the real domain and the complex domain, respectively, and $(\cdot)^\mathsf{T}$ and $(\cdot)^\mathsf{H}$ denote the transpose and the Hermitian of a matrix, respectively. The argument $(t)$ indicates the iteration index of the main algorithm where the subscript $(\cdot)_{(\gamma)}$ indicates the iteration index of the consensus protocols. Finally, the vector $\mathbf{1}$ and $\mathbf{0}$ contain ones and zeros in all entries, respectively. $\mathbf{I}$ is the identity matrix.

\section{Online Distributed Strategy for Eigenvalue Decomposition}\label{sec:disAlg}
A network consisting of $N$ agents is described by the graph $\mathcal{G} = (\mathcal{V},\mathcal{E})$, where $\mathcal{V} = \{1,\ldots,N\}$ denotes the set of nodes (agents) and $\mathcal{E}\subseteq\mathcal{V}\times\mathcal{V}$ defines the set of edges. Throughout the paper, we consider the connected undirected graph $\mathcal{G}$, i.e., the graph $\mathcal{G}$ is characterized by its symmetric adjacency matrix $\mathbf{A} = \left[a_{ij}\right]\in\mathbb{R}^{N\times N}$. The entry $a_{ij}$ is $1$ if $(i,j)\in\mathcal{E}$, i.e., if the $i$-th node has a communication link to the $j$-th node, and $0$ otherwise. The indices of all neighbors of the $i$-th node are collected in the set $\mathcal{N}_i$. Let $d_i = \sum_{j=1}^{N}a_{ij}$ denote the degree of the $i$-th node, which is the number of the elements in the set $\mathcal{N}_i$, then the diagonal matrix $\mathbf{D} = \diag{d_1,d_2,\ldots,d_N}$ represents the degree matrix of $\mathcal{G}$. The corresponding graph Laplacian can be expressed as $\mathbf{L} = \mathbf{D} - \mathbf{A}$. Let $x_i(t)\in\mathbb{C}$ denote the signal of the $i$-th node at the time instant $t$, and then the vector $\mathbf{x}(t) = [x_1(t),x_2(t),\ldots,x_N(t)]^\mathsf{T}\in\mathbb{C}^{N\times 1}$ contains the graph signal of all nodes in the network.

We remark that depending on the particular application scenario, each node can also have access to multiple entries in vector $\mathbf{x}(t)$, and the number of entries in each node can be different. This is for example the case in the decentralized DoA estimation application considered in Section IV-B, where a node has access to the measurements of all sensors in its corresponding subarray. For the simplicity of presentation but without loss of generality, we develop our algorithm based on the case, where each node has a scalar signal $x_i(t)$. The case of multiple signals in each node is however simpler as in this case the associated communications that are required to reach the consensus among the signals of one node are then replaced by the local computation within the respective node.

Our online distributed strategy for eigenvalue decomposition is based on the rank-one modification of a diagonal matrix, where the eigenvalue update is carried out locally, and the rank-one modification is diffused through the network distributively via consensus protocols that will be revised in Section \ref{sec:protocols}. The efficient local update of the rank-one modification using the rational function approximation approach will be explained in the remainder of this section.

\subsection{Rank-One Modification Expression}\label{subsec:rank1}
We address the problem of the online distributed computation of the eigenvalues of a rank-one modification
\begin{equation}\label{equ:problem}
	\mathbf{R}(t) = \mathbf{R}(t-1) + \rho(t) \mathbf{x}(t)\mathbf{x}(t)^\mathsf{T},
\end{equation}
where $\rho(t) \in \{-1,1\}$\footnote{We can always scale equation (\ref{equ:problem}) so that $\rho$ is $1$ or $-1$.}. Denote the eigenvalues $\boldsymbol{\Lambda}(t-1) = \diag{ \lambda_1(t-1),\ldots,\lambda_N(t-1)}$ and corresponding eigenvectors $\mathbf{U}(t-1) = [\mathbf{u}_1(t-1),\ldots,\mathbf{u}_N(t-1)]$ of $\mathbf{R}(t-1)$, which are related as follows\myvspace
\begin{equation}
	\mathbf{U}(t-1)^\mathsf{T}\mathbf{R}(t-1)\mathbf{U}(t-1) = \boldsymbol{\Lambda}(t-1).\myvspace
\end{equation}
Moreover, we assume first that the eigenvalues are distinct and sorted in a descending order as $\lambda_1(t-1)>\cdots>\lambda_N(t-1)$ \footnote{A deflation technique is discussed in Appendix \ref{sec:appDef} to deal with repeated eigenvalues.}.

Multiplying both sides of \eqref{equ:problem} with $\mathbf{U}(t-1)^\mathsf{T}$ and $\mathbf{U}(t-1)$ from the left and the right, respectively, leads to
\begin{equation}\label{equ:rank1modified}
	\begin{aligned}
		\mathbf{U}(t-1)^\mathsf{T} \mathbf{R}(t) \mathbf{U}(t-1) = \boldsymbol{\Lambda}(t-1) + \rho(t) \mathbf{z}(t)\mathbf{z}(t)^\mathsf{T},
	\end{aligned}
\end{equation}
with
\begin{equation}
		\mathbf{z}(t)=\ 
		[z_{1}(t), \ldots, z_{N}(t)]^\mathsf{T} = 
		 \mathbf{U}(t-1)^\mathsf{T}\mathbf{x}(t).\label{equ:disupdate}
\end{equation}
The expression on the right hand side of (\ref{equ:rank1modified}) represents a rank-one modification of a diagonal matrix. The modified eigenvalues and corresponding modified eigenvectors are denoted as $\bar{\boldsymbol{\Lambda}}(t-1)=\diag{\bar{\lambda}_1(t-1),\ldots,\bar{\lambda}_N(t-1)}$ and $\mathbf{V}(t-1) = [\mathbf{v}_1(t-1),\ldots,\mathbf{v}_N(t-1)]$, respectively, which are related as
\begin{equation}
	\label{equ:eigModMat}
	\begin{aligned}
		&\mathbf{V}(t-1)^\mathsf{T}\underbrace{(\boldsymbol{\Lambda}(t-1)+\rho(t)\mathbf{z}(t)\mathbf{z}(t)^\mathsf{T})}_{\mathbf{U}(t-1)^\mathsf{T}\mathbf{R}(t)\mathbf{U}(t-1)}\mathbf{V}(t-1) = \bar{\boldsymbol{\Lambda}}(t-1).
	\end{aligned}
\end{equation}
Furthermore, since 
\begin{equation}
	\mathbf{U}(t)^\mathsf{T}\mathbf{R}(t)\mathbf{U}(t)=\boldsymbol{\Lambda}(t),
\end{equation}
we observe that $\mathbf{R}(t)$ shares the same eigenvalues with the rank-one modified matrix, i.e.,
\begin{equation}\label{equ:eigValUpdate}
	\boldsymbol{\Lambda}(t) = \bar{\boldsymbol{\Lambda}}(t-1),
\end{equation}
and the corresponding eigenvectors can be computed by
\begin{equation}\label{equ:eigVecUpdate}
	\mathbf{U}(t) = \mathbf{U}(t-1)\mathbf{V}(t-1).
	\myvspace
\end{equation}

\subsection{Rational Function Approximation Approach}\label{subsec:ra}
{As described above in Section \ref{subsec:rank1}, the eigenvalue decomposition of a diagonal matrix modified by a rank-one matrix in \eqref{equ:eigModMat} plays a crucial role in the proposed distributed implementation. For notational simplicity, we drop the dependence of the matrix arguments on the time instant $t$. By exploiting the structure of the matrix argument, the efficient implementation of the rank-one modification problem is established by the following theorem \cite{a11}}: 
\begin{theorem}\label{the:rank1}
	Suppose $\boldsymbol{\Lambda} = \diag{\lambda_1,\ldots,\lambda_N}\in\mathbb{R}^{N\times N}$ {where the diagonal entries are distinct and are sorted in descending order, i.e., }  $\lambda_1>\cdots>\lambda_N$. {Further assume} that $\rho\neq 0$ and $\mathbf{z} = \left[z_1, \ldots, z_N\right] \in\mathbb{R}^{N\times 1}$ with $z_i\neq 0$ for all $i = 1, \ldots, N$. If $\mathbf{V} = [\mathbf{v}_1,\cdots,\mathbf{v}_N]\in\mathbb{R}^{N\times N}$ is {an orthogonal matrix} such that
	$$
	\mathbf{V}^\mathsf{T}(\boldsymbol{\Lambda}+\rho \mathbf{z}\mathbf{z}^\mathsf{T})\mathbf{V} = \diag{\bar{\lambda}_1,\ldots,\bar{\lambda}_N},
	$$
	with $\bar{\lambda}_1>\cdots>\bar{\lambda}_N$, then
	\begin{enumerate}
		\item The values in set $\{ \bar{\lambda}_i\}_{i=1}^N$ are the $N$ zeros of the secular function $f(\lambda) = 1 + \rho \mathbf{z}^\mathsf{T}(\boldsymbol{\Lambda}-\lambda\mathbf{I})^{-1}\mathbf{z}$.
		\item The values $\{ \bar{\lambda}_i\}_{i=1}^N$ satisfy the interlacing property, i.e.,\\
		$\bar{\lambda}_1 > \lambda_1 > \bar{\lambda}_2 >  \cdots  >  \bar{\lambda}_N > \lambda_N$, if $\rho > 0$,\\
		$\lambda_1 > \bar{\lambda}_1 > \lambda_2 > \cdots  >  \lambda_N >  \bar{\lambda}_N$, if $\rho < 0$.
		\item The eigenvector $\mathbf{v}_i$ associated with $\bar{\lambda}_i$ is a multiple of $(\boldsymbol{\Lambda} -\bar{\lambda}_i \mathbf{I})^{-1}\mathbf{z}$.$\hfill\blacksquare$
	\end{enumerate}
\end{theorem}

\begin{algorithm}[t]
	\caption{Computing the $k$-th Eigenvalue of Rank-One Modification With Rational Function Approximation, $\mathtt{RA}_k(\cdot)$}
	\label{alg:ra}
	\begin{algorithmic}[1]
		\STATE \textbf{Initialization}: Iteration index $\tau = 0$, scalar $\rho$, vector $\mathbf{z}$, tolerance $\xi$, arbitrary starting point $\lambda^{(\tau)}\in(\lambda_{k},\lambda_{k-1})$
		\REPEAT
		\STATE Find the parameters $p$ and $q$ such that
		\begin{equation}
			\tilde{\psi}_{k-1}(\lambda^{(\tau)}) = \psi_{k-1}(\lambda^{(\tau)})\text{ and } \tilde{\psi}_{k-1}^\prime(\lambda^{(\tau)}) = \psi_{k-1}^\prime(\lambda^{(\tau)}).
		\end{equation}
		\STATE Find the parameters $r$ and $s$ such that
		\begin{equation}
			\tilde{\phi}_{k}(\lambda^{(\tau)}) = \phi_{k}(\lambda^{(\tau)})\text{ and }\tilde{\phi}_{k}^\prime(\lambda^{(\tau)}) = \phi_{k}^\prime(\lambda^{(\tau)}).
		\end{equation}
		\STATE Find $\lambda^{(\tau + 1)}\in(\lambda_{k},\lambda_{k-1})$ which satisfies\begin{equation}
			-\tilde{\psi}_{k-1}(\lambda^{(\tau+1)}) = 1 + \tilde{\phi}_k(\lambda^{(\tau+1)})
		\end{equation}
		\STATE $\tau \gets \tau + 1$
		\UNTIL{$|\lambda^{(\tau+1)} - \lambda^{(\tau)}| < \xi$}
		\RETURN $\bar{\lambda}_k = \lambda^{(\tau + 1)}$, $\mathbf{v}_k = \frac{(\boldsymbol{\Lambda}-\bar{\lambda}_k\mathbf{I})^{-1}\mathbf{z}}{\left\|(\boldsymbol{\Lambda}-\bar{\lambda}_k\mathbf{I})^{-1}\mathbf{z}\right\|_2}$
	\end{algorithmic}	
\end{algorithm}

There is no loss in generality in assuming that $\rho > 0$, otherwise, we can replace $\lambda_i$ by $-\lambda_{N-i+1}$ and $\rho$ by $-\rho$ \cite{a11}.
According to the \textit{Theorem \ref{the:rank1}.1}, the eigenvalues $\bar{\lambda}_k$ of the matrix $\boldsymbol{\Lambda} + \rho \mathbf{z}\mathbf{z}^\mathsf{T}$ can be computed by solving $f(\lambda) = 0$, i.e.,
\myvspace
\begin{equation}
	f(\lambda) = 1 + \rho\sum_{n = 1}^{N}\frac{z_{i}^2}{\lambda_i-\lambda} = 0.
\end{equation}
Based on the interlacing property of the eigenvalues, for the $k$-th eigenvalue $\bar{\lambda}_k\in(\lambda_{k},\lambda_{k-1})$ with $\lambda_0 = \lambda_1 + \rho \mathbf{z}^\mathsf{T}\mathbf{z}$ \cite{a90}, we can rearrange the equation as
\begin{equation}\label{equ:ra}
	\begin{aligned}
		-\psi_{k-1}(\lambda)  & = 1 + \phi_k(\lambda) 
	\end{aligned}
\end{equation}	
where
\begin{equation}
\label{equ:trueFunc}
		\psi_{k-1}(\lambda)  = \rho \sum_{i = 1}^{k-1}\frac{z_{i}^2}{\lambda_i-\lambda}\ \text{ and }\  \phi_k(\lambda)  = \rho \sum_{i = k}^{N}\frac{z_{i}^2}{\lambda_i-\lambda}.
\end{equation}
Since both functions $\psi_{k-1}(\lambda)$ and $\phi_k(\lambda)$ are sums of rational functions, it is natural to approximate them with simple rational functions \cite{a11} as
\begin{equation}
\label{equ:approxFunc}
\tilde{\psi}_{k-1}(\lambda) =\ p + \frac{q}{\lambda_{k-1} - \lambda}\ \text{ and }\  \tilde{\phi}_{k}(\lambda) =\ r + \frac{s}{\lambda_{k} - \lambda}.
\end{equation}
{In \eqref{equ:approxFunc}, the parameters $p$, $q$, $r$ and $s$ are chosen such that, at the given iterate $\lambda^{(\tau)}$, the rational approximants $\tilde{\psi}_{k-1}(\lambda)$ and $\tilde{\phi}_k(\lambda)$ in \eqref{equ:approxFunc} coincide with the true rational functions $\psi_{k}(\lambda)$ and $\psi_{k-1}(\lambda)$ in \eqref{equ:trueFunc} up to the first derivative, respectively. The next iterate $\lambda^{(\tau+1)}\in\left(\lambda_k, \lambda_{k-1}\right)$ is obtained from a solution of the following equation
\begin{equation}
	-\tilde{\psi}_{k-1}(\lambda)  = 1 + \tilde{\phi}_k(\lambda).\myvspace
\end{equation}
}For the special case where $k = 1$, function $\psi_{k-1}(\lambda)$ is approximated as $\tilde{\psi}_0(\lambda) = 0$. The rational function approximation algorithm is summarized in Algorithm \ref{alg:ra} \cite{a22}.

\subsection{Online Distributed Eigenvalue Decomposition Protocol}\label{subsec:onlineAlg}
In this section we describe how the eigendecomposition of a symmetric matrix with a rank-one modification, i.e., matrix $\mathbf{R}(t)$ in \eqref{equ:problem}, is computed in the network based on the distributed evaluation of the linear equation in \eqref{equ:disupdate} using a distributed consensus protocol. In contrast to the \emph{centralized scheme}, where the measurements of each node are communicated to all nodes in the network (or to a central fusion center), the direct communication of the measurements $\mathbf{x}(t)$ and the explicit computation of $\mathbf{R}(t)$ are avoided in our \emph{decentralized scheme}. Before we revise different distributed consensus protocols in Section \ref{sec:protocols}, we outline the general procedure to distribute the algorithm as summarized in Algorithm \ref{alg:onlineDis}.

In Algorithm \ref{alg:onlineDis}, we assume that each node maintains one row of the current and the past eigenvector matrices $\mathbf{U}(t)$ and $\mathbf{U}(t-1)$, respectively. Specifically, the $i$-th node locally stores vectors $\tilde{\mathbf{u}}_i^\mathsf{T}(t) = \mathbf{e}_i^\mathsf{T} \mathbf{U}(t) \in \mathbb{C}^{N\times 1}$ and $\tilde{\mathbf{u}}_i^\mathsf{T}(t-1) = \mathbf{e}_i^\mathsf{T} \mathbf{U}(t-1) \in \mathbb{C}^{N\times 1}$. Furthermore, each node maintains instances, i.e., local copies, of the current and past diagonal matrices of eigenvalues ${\boldsymbol \Lambda}(t)$ and ${\boldsymbol \Lambda}(t-1)$, respectively, as well as the auxiliary vectors $\mathbf{v}_k(t-1)$ and $\mathbf{z}(t)$. For the $i$-th node the local instances of the auxiliary vectors are denoted as $\mathbf{v}_{i,k}(t-1)$ and $\mathbf{z}_i(t)$\footnote{The indices $k$ and $t$ for auxiliary variables $\mathbf{z}_i(t)$ and $\mathbf{v}_{i,k}(t-1)$ are kept for the consistent presentation of the algorithm. In practical implementations, the storage for each auxiliary variable can be reused.}. Then following \eqref{equ:disupdate}, the $k$-th entry of the update vector $\mathbf{z}_i(t) = [z_{i,1}(t),z_{i,2}(t),\ldots,z_{i,N}(t)]^\mathsf{T}$ is distributively updated by
\begin{equation}\label{equ:zupdate}
	z_{i,k}(t) = \tilde{\mathbf{u}}_k^\mathsf{T}(t-1)\mathbf{x}(t) = \sum_{j = 1}^{N}u_{jk}(t-1)x_j(t).
\end{equation}
According to (\ref{equ:eigVecUpdate}), the entries of the $i$-th row of $\mathbf{U}(t)$ are locally updated by 
\begin{equation}\label{equ:eigVecUpdateLocal}
	\tilde{u}_{i,k}^\mathsf{T}(t) = \tilde{\mathbf{u}}_{i}^\mathsf{T}(t-1)\mathbf{v}_{i,k}(t-1),\quad \text{for } k = 1, \ldots, N. 
\end{equation}

\begin{algorithm}[t]
	\caption{Online Distributed Eigenvalue Decomposition over Network}
	\label{alg:onlineDis}
	\begin{algorithmic}[1]
		\STATE \textbf{Initialization}: In the $i$-th node $\boldsymbol{\Lambda}_i(0) = \mathbf{0}, \tilde{\mathbf{u}}_i^\mathsf{T}(0) = \mathbf{e}_i^\mathsf{T}$, $\rho(t)$, $t = 1, \forall i \in\mathcal{V}$ \label{alg:onlineDis:init}
		\WHILE{$\mathbf{x}(t)$ observed} \label{alg:onlineDis:x}
		\STATE \underline{Network Communication}
		\FOR{each node $i\in\mathcal{V}$}
		\FOR{each entry $k = 1,\ldots,N$}
		\STATE $z_{i,k}(t) = \mathtt{NC}_k(\tilde{\mathbf{u}}_k(t-1)^\mathsf{T}\mathbf{x}(t))$ (Eq. \eqref{equ:zupdate}) \label{alg:onlineDis:z}
		\ENDFOR
		\ENDFOR
		\STATE \underline{Local Computation}
		\FORALLN{node $i\in \mathcal{V}$}
		\FOR{each entry $k = 1,\ldots,N$}
		\STATE $[\bar{\lambda}_{i,k}(t-1),\mathbf{v}_{i,k}(t-1)] = \mathtt{RA}_k(\boldsymbol{\Lambda}_i(t-1) + \rho(t)\mathbf{z}_i(t)\mathbf{z}_i(t)^\mathsf{T})$ (Alg. \ref{alg:ra})\label{alg:onlineDis:RA}
		\STATE Update $\lambda_{i,k}(t) = \bar{\lambda}_{i,k}(t-1)$ (Eq.~(\ref{equ:eigValUpdate}))  \label{alg:onlineDis:eigVal}
		\STATE Update $\tilde{u}_{i,k}^\mathsf{T}(t)= \tilde{\mathbf{u}}_i^\mathsf{T}(t-1)\mathbf{v}_{i,k}(t-1)$  (Eq.~(\ref{equ:eigVecUpdateLocal})) \label{alg:onlineDis:eigVec}
		\ENDFOR
		\ENDFAN
		\ENDWHILE
	\end{algorithmic}	
\end{algorithm}
	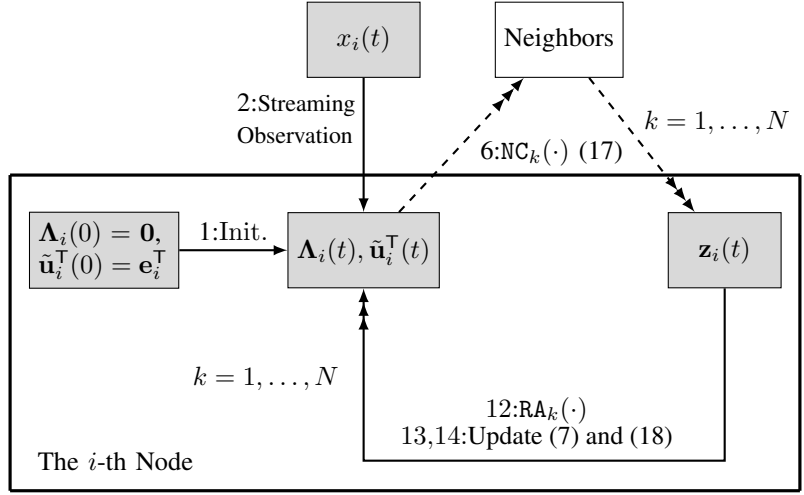
\begin{figure}[t]
	\centering
	\begin{tikzpicture}[scale=1]
		\node[rectangle,draw,fill=black!15!white, minimum width=1.5cm, minimum height=1cm,text width=1.75cm](0) at (-3.25,0) {{$\boldsymbol{\Lambda}_i(0) = \mathbf{0}$, $\tilde{\mathbf{u}}_i^\mathsf{T}(0)=\mathbf{e}_i^\mathsf{T}$}};
		
		\node[rectangle,draw,fill=black!15!white, minimum width=1.3cm, minimum height=1cm](1) at (0.2,0) {{$\boldsymbol{\Lambda}_i(t),\tilde{\mathbf{u}}_i^\mathsf{T}(t)$}};
		
		\node[rectangle,draw,fill=black!15!white, minimum width=1.5cm, minimum height=1cm](2) at (5,0) {{$\mathbf{z}_i(t)$}};
		\node[rectangle,draw,fill=black!15!white, minimum width=1.5cm, minimum height=1cm](x) at (0.2,2.8) {{$x_i(t)$}};
		\node[rectangle,draw,fill=none, minimum width=1.5cm, minimum height=1cm](network) at (2.8,2.8) {{Neighbors}};
		
		\draw[very thick] (-4.5,1) -- (6,1);
		\draw[very thick] (6,1) -- (6,-3.2);
		\draw[very thick] (-4.5,-3.2) -- (6,-3.2);
		\draw[very thick] (-4.5,1) -- (-4.5,-3.2);
		\node (nodei) at (-3.1,-2.8) {The $i$-th Node};
		\node (nc) at (2.7,1.3) {{\ref{alg:onlineDis:z}:}$\mathtt{NC}_k(\cdot)$ \eqref{equ:zupdate}};
		\node (k1Nnc) at (4.9, 1.7) {$k = 1,\ldots,N$};
		
		\draw[-latex,thick] (0) -- (1) node [pos=0.5, above] {{\ref{alg:onlineDis:init}:}$\mathrm{Init.}$};
		\draw[-{latex}{latex}{latex},thick,dashed] (1) -- (network);
		\draw[-{latex}{latex}{latex},thick,dashed] (network) -- (2) ;
		\draw[-{latex}{latex}{latex},thick] (2) -- (5,-2.8) -- (0.2,-2.8) -- (1);
		\draw[-latex,thick] (x) -- (1) node [pos=0.3, left, text width=1.55cm] {{\ref{alg:onlineDis:x}:}\small{Streaming Observation}};
		\node(ra) at (2.5,-2.3) {\Large{$\substack{\ref{alg:onlineDis:RA}:\mathtt{RA}_k(\cdot)\\\ref{alg:onlineDis:eigVal},\ref{alg:onlineDis:eigVec}:\text{Update } \eqref{equ:eigValUpdate} \text{ and } \eqref{equ:eigVecUpdateLocal}}$}};
		\node (k1Nra) at (-1.1, -1.7) {$k = 1,\ldots,N$};
	\end{tikzpicture}
	\caption{Working flow in the $i$-th node, where the $\mathtt{NC}_k(\cdot)$ step, indicated with dashed arrows, requires the network collaboration. Arrows with multiple heads indicate that associated operations need to be carried out for $k=1,\ldots,N$ in each iteration. Indices of the operations are line indices in Algorithm \ref{alg:onlineDis}.}
	\label{fig:workflow}
\end{figure}

The update procedure of Algorithm \ref{alg:onlineDis} is illustrated as a flow chart in Figure \ref{fig:workflow}, where the local information, the local
updates, and the network communication are indicated by gray
boxes, solid arrows, and dashed arrows, respectively. The $i$-th node is initialized with $\boldsymbol{\Lambda}_i(0) = \mathbf{0}$ and $\tilde{\mathbf{u}}_i^\mathsf{T}(0) = \mathbf{e}_i^\mathsf{T}$.
At the time instant $t$, the $i$-th node contributes a new sample $x_i(t)$. Firstly, the required Network Communication (NC) is carried out, where 
the $k$-th entry of $\mathbf{z}_i(t)$ in the $i$-th node is computed distributively throughout the network according to (\ref{equ:zupdate}) by running any distributed consensus protocol denoted by $\mathtt{NC}_k(\cdot)$. Different distributed consensus algorithms can be applied in this step, which will be revised in Section \ref{sec:protocols}. Then the local update is performed by applying the rational function approximation locally in each node, which is possible since the local instances $\boldsymbol{\Lambda}_i(t-1), \rho(t)$, and $\mathbf{z}_i(t)$ are accessible to the $i$-th node. Instead of performing the update \eqref{equ:eigVecUpdateLocal} fully parallelized, if the storage capacity is not available the rational function approximation can be carried out partially parallelized or fully sequentially, such that in the latter case the storage requirement for the auxiliary vectors $\mathbf{v}_{i,k}(t-1)$ is collapsed to one vector of size $N$.
\begin{remark}
	Our proposed decentralized eigendecomposition scheme has the following advantages over the centralized eigendecomposition implementation. First, the observations are naturally distributed in the network and not available to each node (or the central processing node). Second, the decentralized implementation is simple and robust to, e.g., node failure, as centralized processing is associated with the requirement to route the data of all nodes to a centralized processor, where the failure in, e.g., the central processor leads to a complete breakdown of the application. Third, the decentralized scheme is more suitable for scalability than the centralized scheme with respect to the memory/storage requirements, which would be drastically increased in a centralized processing scheme. To be precise, in the proposed scheme the storage requirement of each node is $6N$ real floating point values, which is linear in the size of the network $N$, for the local vector instances $\mathbf{z}_i(t)$, the auxiliary variable $\mathbf{v}_{i,k}(t-1)$, the $i$-th row of matrices $\mathbf{U}(t)$ and $\mathbf{U}(t-1)$, i.e., $\tilde{\mathbf{u}}_i^\mathsf{T}(t)$ and $\tilde{\mathbf{u}}_i^\mathsf{T}(t-1)$, respectively, as well as the eigenvalues on the diagonal of matrix $\boldsymbol{\Lambda}_i(t)$ and $\boldsymbol{\Lambda}_i(t-1)$. In contrast, in the scheme where the entire vector $\mathbf{x}(t)$ is communicated to each node (or the central server), each node needs to store the entire matrix $\mathbf{U}(t)$ and $\mathbf{U}(t-1)$ instead of just one row, along with the remaining vectors, resulting in a total storage requirement of $2N^2 + 4N$ real floating point values, which increase quadratically in $N$. Finally, even thought negligible, the computation cost in each node is reduced, since the $i$-th node only updates its corresponding $i$-th row of $\mathbf{U}(t)$ instead of the whole matrix in a centralized scheme.
\end{remark}

\begin{remark}
	{In the first $N-1$ samples with $t = 1, \ldots, N-1$, the matrix $\mathbf{R}(t)$ has zero eigenvalues with multiplicities, which violates the assumptions in \textit{Theorem~\ref{the:rank1}}}. Thus, an extra deflation step is required to remove the multiplicity and to obtain a rank-one modification with a smaller size. Following the deflation technique discussed in \cite{a11}, where the potential case that $\mathbf{z}(t)$ contains zero components can also be deflated, we illustrate a deflation method using the Householder transformation in Appendix \ref{sec:appDef}. Notice that no extra network communication is required since the deflation is done locally in each node. Moreover, by replacing transpose with hermitian \textit{Theorem~\ref{the:rank1}} provides similar result if the rank-one update $\mathbf{z}(t)$ is complex.
\end{remark}

\section{Consensus Protocols}\label{sec:protocols}
The decentralized computation of the entries of the rank-one update vector plays a crucial role in our decentralized algorithm, and we remark that the decentralized computation of the expression in (\ref{equ:zupdate}), i.e., $\mathtt{NC}_k(\cdot)$ for the $k$-th entry of $\mathbf{z}(t)$, can be carried out with various methods, e.g., averaging consensus protocols, linear graph filters, and nonlinear graph filters \cite{a105,a106}. In the following, we revise four different approaches that can be utilized for the proposed online decentralized eigendecomposition algorithm. This includes three conventional averaging consensus protocols and a low-pass graph filter approach, where we propose to employ the normalized adjacency matrix as the shift operator to ensure the stability of the graph filter.
\subsection{Push-Sum (PS) Consensus Algorithm}\label{subsec:ps}
One prominent candidate for computing the weighted sum in (\ref{equ:zupdate}) distributively is the PS consensus algorithm, which was first introduced and analyzed in \cite{a88}, and its convergence is proven in \cite{a89} for arbitrary graphs based on weak ergodicity arguments. Its principle is provided as follows. %

{Assume that the vector $\mathbf{y} = \left[y_1, \ldots, y_N\right]^\mathsf{T}$ contains the graph signal of $N$ nodes whose average needs to be computed  distributively over the network}. We introduce a column stochastic matrix $\mathbf{P}$, i.e., $\mathbf{1}^\mathsf{T}\mathbf{P}=\mathbf{1}^\mathsf{T}$, where $p_{ji} = 0$ if there is no direct edge between the $i$-th and the $j$-th node. A simple and sufficient example of the matrix $\mathbf{P}$ is \myvspace
\begin{equation}
	p_{ji} = \begin{cases}
		1/d_i, \quad &(i,j)\in\mathcal{E},\\
		0,\quad &\text{otherwise}.
	\end{cases}\myvspace
\end{equation}

{In order to perform the averaging operation distributively, the PS consensus algorithm further assumes that,} at a given consensus iteration $\gamma$, the $i$-th node maintains a set consisting of two values: a cumulative estimate of the sum $s_{i(\gamma)}$ and a weight $w_{i(\gamma)}$ for $i = 1,\ldots,N$. {The vector of sums $\mathbf{s}_{(\gamma)} = \left[s_{1(\gamma)}, \ldots, s_{N(\gamma)}\right]^\mathsf{T}\in\mathbb{R}^{N}$ and the vector of weights $\mathbf{w} = \left[w_{1(\gamma)}, \ldots, w_{N(\gamma)}\right]^\mathsf{T}\in\mathbb{R}^{N}$} are initialized as, for example, \myvspace
\begin{equation}
	\mathbf{s}_{(0)} = \mathbf{y}\quad \text{and}\quad \mathbf{w}_{(0)} = \mathbf{1},\myvspace
\end{equation}
respectively, which are locally available at the nodes. {The PS algorithm consists of two steps, which are iteratively performed in all nodes of the network until convergence. At the $\gamma$-th consensus iteration,} based on the chosen column stochastic matrix $\mathbf{P}$, the $i$-th node first splits its total sum $s_{i(\gamma)}$ and weight $w_{i(\gamma)}$ into shares and sends to its neighboring $j$-th node the corresponding share $\mathbb{S}_{i\rightarrow j(\gamma)} = \left\lbrace p_{ji}s_{i(\gamma)},p_{ji}w_{i(\gamma)}\right\rbrace, \forall j \in\mathcal{N}_i$. {Then, each node updates its own sum and weight by summing up all the shares received from its adjacent nodes.} The above mentioned process is summarized in vector form as \myvspace
\begin{equation}
\label{equ:updInfo}
	\mathbf{s}_{(\gamma)} = \mathbf{P}\mathbf{s}_{(\gamma-1)}\quad \text{and}\quad \mathbf{w}_{(\gamma)} = \mathbf{P}\mathbf{w}_{(\gamma-1)}.\myvspace
\end{equation}
{Given the estimated sum and weight in \eqref{equ:updInfo},} the estimated average is calculated at each node as\myvspace
\begin{equation}
	\hat{\mathbf{z}}_{(\gamma)} = \mathbf{s}_{(\gamma)}\oslash\mathbf{w}_{(\gamma)},\myvspace
\end{equation}
where $\oslash$ is Hadamard, i.e., elementwise division. It can be shown that the PS algorithm converges at each node to the same average value, i.e.,\myvspace
\begin{equation}
	\lim_{\gamma\rightarrow\infty}\hat{z}_{i(\gamma)} = \lim_{\gamma\rightarrow\infty}\frac{s_{i(\gamma)}}{w_{i(\gamma)}} = \frac{\mathbf{1}^\mathsf{T}\mathbf{y}}{N}, \ \forall i \in\mathcal{V}.\myvspace\footnote{If only one node starts with weight $1$, where all the others start with weight $0$, then the value computed at the nodes converges to the sum $\mathbf{1}^\mathsf{T}\mathbf{y}$, instead of their average. Furthermore, if every node starts with value $s_{i(\gamma)} = 1$, the size of the network can be determined distributively.}
\end{equation}

The PS consensus algorithm is summarized in Algorithm \ref{alg:ps}.\myvspace
\begin{algorithm}[t]
	\caption{Push-Sum Consensus Algorithm Performed at the $i$-th Node for the $k$-th Entry of $\mathbf{z}(t)$, $\mathtt{PS}_k(\cdot)$}
	\label{alg:ps}
	\begin{algorithmic}[1]
		\STATE \textbf{Initialization}: Iteration index $\gamma = 0$, maximum {number of consensus} iterations $\Gamma$, $p_{ji}$, initialize $s_{i(0)} = u_{ik}(t)x_i(t)$ (Eq.~\eqref{equ:zupdate}) and $w_{i(0)} = 1$.%
		\WHILE{$\gamma\leq\Gamma$}
		\STATE \textbf{\textit{Push step}}: Send the shares $\mathbb{S}_{i\rightarrow j(\gamma)}$ to all adjacent nodes $j\in\mathcal{V}$ with $(i,j)\in\mathcal{E}$.
		\STATE \textbf{\textit{Sum step}}: Sum the shares $\mathbb{S}_{j\rightarrow i(\gamma)}$ obtained from all adjacent nodes $j \in\mathcal{V}$ with $(j,i)\in\mathcal{E}$.
		\STATE $\gamma \gets \gamma + 1$
		\ENDWHILE
		\RETURN{$z_k(t)= Ns_{i(\gamma+1)}/w_{i(\gamma+1)}$}
	\end{algorithmic}	
\end{algorithm}

\subsection{Average Consensus (AC) Algorithm}\label{subsec:fac}
Apart from the PS algorithm, the AC algorithm, first introduced and analyzed in \cite{a108}, can also be applied for the distributed computation of a weighted sum. %
The principle of the AC algorithm is provided as follows.

Based on the graph Laplacian $\mathbf{L}$, we introduce a commonly used update matrix 
\begin{equation}
	\mathbf{W}  = \mathbf{I} - \epsilon\mathbf{L},
\end{equation} 
where $\epsilon$ is the step size, which must satisfy $\epsilon<\frac{2}{\lambda_\mathtt{max}(\mathbf{L})}$\cite{a107}. The update matrix in this case is doubly stochastic, i.e., $	\mathbf{W}\mathbf{1} = \mathbf{1}$ and $\mathbf{1}^\intercal\mathbf{W} = \mathbf{1}^\intercal$, 
and the step size can be chosen according to the maximum-degree weight, i.e.,
\begin{equation}\label{equ:stepsizemd}
	\epsilon^\text{md} = \frac{1}{d_\text{max}},
\end{equation}
where $d_{\text{max}}$ indicates the maximum degree of the nodes.

For the $i$-th node, the AC algorithm is initialized with $s_{i(0)} =~y_i$ and at the $\gamma$-th consensus iteration, the $i$-th node updates its own values based on the disagreement with the neighboring nodes with the step size $\epsilon$. This is expressed as
\begin{equation}\label{equ:ac}
	s_{i(\gamma)} = s_{i(\gamma-1)} - \epsilon\sum_{j\in\mathcal{N}_i}(s_{i(\gamma-1)}-s_{j(\gamma-1)}),
\end{equation}
which can also be compactly written as
\begin{equation}
	\mathbf{s}_{(\gamma)} = \mathbf{W}\mathbf{s}_{(\gamma-1)}.
\end{equation}

The AC algorithm is summarized in Algorithm \ref{alg:ac}.
\begin{algorithm}[t]
	\caption{Average Consensus Protocol Performed at the $i$-th Node for the $k$-th Entry of $\mathbf{z}(t)$, $\mathtt{AC}_k(\cdot)$}
	\label{alg:ac}
	\begin{algorithmic}[1]
		\STATE \textbf{Initialization}: Iteration index $\gamma = 0$, maximum {number of consensus} iterations $\Gamma$, step size $\epsilon^\text{md}$ (Eq.~\eqref{equ:stepsizemd}), initialize $s_{i(0)} = u_{ik}(t)x_i(t)$ (Eq.~\eqref{equ:zupdate}).
		\WHILE{$\gamma\leq\Gamma$}
		\STATE Update the estimate $s_{i(\gamma)}$ (Eq.~\eqref{equ:ac})
		\STATE $\gamma \leftarrow \gamma + 1$
		\ENDWHILE
		\RETURN{$z_k(t)= Ns_{i(\gamma+1)}$}
	\end{algorithmic}	
\end{algorithm}

\subsection{Finite-Time Average Consensus (ftAC) Algorithm}\label{subsec:ftac}
The AC algorithm may suffer from slow convergence. To guarantee the exact convergence after finite number of iterations, the ftAC algorithm has been proposed where the knowledge of the graph Laplacian $\mathbf{L}$ plays a central role \cite{a109,a111}.

For the ftAC algorithm, the update matrix $\mathbf{W}$ is no longer constant but adapts with consensus iterations. The sequence of update matrices $\{\mathbf{W}_{(0)},\mathbf{W}_{(1)},\ldots\}$ is not unique, and one choice of the sequence is based on the eigenvalues of the graph Laplacian. Without loss of generality, suppose the distinct eigenvalues of $\mathbf{L}$ are $\lambda_1^{\star},\ldots,\lambda_R^{\star} = 0$, where $R\leq N$, then the update matrices are chosen as \cite{a109,a111}
\begin{equation}\label{equ:ftAC}
	\mathbf{W}_{(\gamma)} = \mathbf{I} - \frac{1}{\lambda_{\gamma+1}^\star}\mathbf{L},\quad \gamma = 0,\ldots,R-2.
\end{equation}
The choice of update matrices shown in \eqref{equ:ftAC} can be considered as a special case that is used in the conventional AC algorithm with adaptive step sizes chosen as
\begin{equation}\label{equ:stepsizeftAC}
	\epsilon_{(\gamma)}^\star = \frac{1}{\lambda^\star_{\gamma+1}},\quad \gamma = 0,\ldots,R-2.
\end{equation}
By choosing the aforementioned sequence of update matrices, the ftAC algorithm guarantees the exact convergence after $R-1$ iterations, i.e., the number of the distinct nonzero eigenvalues of the graph Laplacian $\mathbf{L}$. 

The ftAC algorithm is summarized in Algorithm \ref{alg:ftac}.
\begin{algorithm}[t]
	\caption{Finite-time Average Consensus Protocol Performed at the $i$-th Node for the $k$-th Entry of $\mathbf{z}(t)$, $\mathtt{ftAC}_k(\cdot)$}
	\label{alg:ftac}
	\begin{algorithmic}[1]
		\STATE \textbf{Initialization}: Iteration index $\gamma = 0$, distinct eigenvalues of the graph Laplacian: $\lambda_1^{\star},\ldots,\lambda_R^{\star} = 0$, initialize $s_{i(0)} = u_{ik}(t)x_i(t)$ (Eq.~\eqref{equ:zupdate}).
		\WHILE{$\gamma\leq R-2$}
		\STATE Update the step size $\epsilon_{(\gamma)}^\star$ (Eq.~\eqref{equ:stepsizeftAC})
		\STATE Update the estimate $s_{i(\gamma)}$ (Eq.~\eqref{equ:ac})
		\STATE $\gamma \leftarrow \gamma + 1$
		\ENDWHILE
		\RETURN{$z_k(t)= Ns_{i(\gamma+1)}$}
	\end{algorithmic}	
\end{algorithm}

\subsection{Graph Filter Method}\label{subsec:gfilterConsensus} 
Instead of average consensus protocols, an alternative is to use low-pass graph filters. Denote the graph shift operator as $\mathbf{S} \in \mathbb{R}^N$ to describe the interactions between neighboring nodes, and the eigenvalues of the shift operator are also known as the graph frequencies. 
A linear shift-invariant graph filter $\widehat{\mathbf{H}}$ can be implemented in a distributed fashion and the output $\mathbf{y}$ of the graph filter is related to the input $\mathbf{x}$ as
\begin{equation}\label{equ:graphFilterxy}
	\mathbf{y} = \widehat{\mathbf{H}}\mathbf{x},
\end{equation}
where
\begin{equation}\label{equ:graphFilterH}
	\widehat{\mathbf{H}} = h_0\mathbf{S}^0 + h_1\mathbf{S}^1 + h_2\mathbf{S}^2 + \cdots = \sum_{m = 0}^{K}h_m\mathbf{S}^m,
\end{equation}
and $h_m \in \mathbb{R},\ m = 0, \ldots, K,$ are polynomial coefficients for filter order $K$. Then the frequency response of the graph filter at the frequency $\lambda$ is 
\begin{equation}\label{equ:graphFilterResponse}
	\hat{h}(\lambda) = \sum_{m = 0}^{K}h_m\lambda^m.
\end{equation}
If the structure of the graph is unknown or the direct eigendecomposition of the shift operator is not practical, a universal design, i.e., graph independent filter design, is carried out, where only the range of the eigenvalues is required, and the high computation cost of the direct eigendecomposition is avoided. Nevertheless, graph dependent filters reduce the filter length and therefore the communication overhead in the consensus procedure \cite{a112}, which however, requires the knowledge of the graph spectrum.

To carry out the consensus operation, i.e., the distributed computation of the weighted average in (\ref{equ:zupdate}), a low-pass graph filter is required, where only the graph signal components associated with the low frequencies are preserved. For example, when the graph Laplacian is applied as the graph shift operator, i.e., $\mathbf{S} = \mathbf{L}$, the graph signal components associated with high frequencies, i.e., $\lambda > 0$, are suppressed, while those associated with the low frequency, i.e., $\lambda = 0$, are preserved. We remark that for more general variations of the shift operator the ordering of the graph frequencies from low to high may differ from the magnitudes of the eigenvalues of the shift operator. Instead, the ordering of the graph frequencies is defined, e.g., according to the total variation of the associated eigenvectors or their linear transformations \cite{a132}. The polynomial coefficients of the graph filter can be computed, e.g., by the polynomial fitting, in the case of the graph dependent filter design, at the known graph frequencies or, alternatively, in the graph independent filter design, at candidate frequencies on a sampling grid.%

Nevertheless, due to the fact that the multiplication with high power of the shift operator results in the amplification of intermediate graph signals and potential round off errors in the filter in large networks, we propose to use the normalized adjacency matrix $\bar{\mathbf{L}} = \mathbf{D}^{-\frac{1}{2}}\mathbf{A}\mathbf{D}^{-\frac{1}{2}}= \mathbf{I} - \mathbf{D}^{-\frac{1}{2}}\mathbf{L}\mathbf{D}^{-\frac{1}{2}}$ as the shift operator, since its eigenvalues 
lie between $-1$ and $1$, and thus the amplification of intermediate graph signals is reduced, and the round off errors can be avoided. For the concern of the decentralized implementation, the mapping from $\mathbf{L}$ to $\bar{\mathbf{L}}$ is easily carried out in each node locally as $\mathbf{D}$ is diagonal (cf. Section \ref{subsec:gfilter} and Appendix \ref{sec:appShiftOperator} for further discussion).

We remark that all aforementioned protocols can be applied in the network communication step in our proposed online distributed eigenvalue decomposition algorithm as stated in Algorithm \ref{alg:onlineDis}. Moreover, different protocols provide different benefits. In particular, while the AC algorithm is favorable with its simple implementation, its application is limited to undirected graphs. On directed graphs, the PS algorithm can be applied, which maintains two variables in each iteration. While the AC algorithm and the PS algorithm convergence asymptotically, the ftAC algorithm and the graph filter approach can reach the consensus in finite number of iterations, i.e., in the number of distinct non-zero eigenvalues of the graph shift operator $R-1$, and in the filter order $K$, respectively. Furthermore, the local storage and computation requirements of different protocols also vary. The comparison between different consensus protocols is summarized in Table \ref{tab:prot}.

\begin{table*}[t]
	\caption{Summation of different consensus protocols.}
	\centering
	\begin{tblr}{c|c|c|c|c}
		Protocol 	& Graph 		& Convergence speed	 & Local Storage& Extra Local Computation\\
		\hline
		PS		 	& (un)directed& asymptotic ($\Gamma \to \infty$)& $s_{i(\gamma)},w_{i(\gamma)}$	& $s_{i(\gamma)}/w_{i(\gamma)}$  \\
		\hline
		AC       	& undirected	 & asymptotic ($\Gamma \to \infty$)& $s_{i(\gamma)},\epsilon^\text{md}$	& $\epsilon^\text{md}$\\
		\hline
		ftAC	 	& undirected  & finite time ($R-1$) & $s_{i(\gamma)},\epsilon_{(\gamma)}^\star$ & $\lambda_0,\ldots,\lambda_{R-1},\epsilon_{(\gamma)}^\star$\\
		\hline
		Graph Filter& (un)directed& finite time ($K$) & $h_0,\ldots,h_K$ & $h_0,\ldots,h_K$	\\
		\hline
	\end{tblr}	
	\label{tab:prot}
\end{table*}

\section{Application Examples}\label{sec:simulation}
In this section, we study important application examples that rely on the distributed eigenvalue computation. Throughout this section, the numerical precision for the rational function approximation according to Algorithm \ref{alg:ra} is set as $\xi = 10^{-12}$. We use the PS algorithm as the consensus protocol if not specified otherwise.

\subsection{Distributed Sample Covariance Spectrum Estimation and Subspace Tracking}\label{subsec:cov}
One prominent application example of our distributed algorithm is the distributed tracking of the eigenvalue decomposition of the moving sample covariance matrix. In this application, $x_i(t)$ is the observation obtained at the $i$-th node, and the exponential weighted moving sample covariance matrix of the graph signals is given as
\begin{equation}\label{equ:sampleCov}
	\mathbf{R}(t) = \alpha\mathbf{R}(t-1) + (1- \alpha)\mathbf{x}(t)\mathbf{x}(t)^\mathsf{H},
\end{equation}
where $0 < \alpha < 1$ is the forgetting factor. This is exactly a rank-one modification problem stated as (\ref{equ:problem}) discussed in Section \ref{sec:sysMod}. Another commonly used variant of the moving sample covariance matrix is obtained using a sliding window
\begin{equation}\label{equ:sampleCov2}
	\mathbf{R}_\beta(t) = \mathbf{R}_\beta(t-1) + \mathbf{x}(t)\mathbf{x}(t)^\mathsf{H} - \mathbf{x}(t-\beta)\mathbf{x}(t-\beta)^\mathsf{H},
\end{equation}
where $\beta$ is the length of the sliding window. Note that the scalar $\rho$ in (\ref{equ:problem}) can be either positive or negative, and the spectrum computation and subspace tracking of Equation (\ref{equ:sampleCov2}) can be obtained from Algorithm \ref{alg:onlineDis} when we treat update \eqref{equ:sampleCov2} as two consecutive rank-one modifications.

\begin{figure}[t]
	\centering
	\input{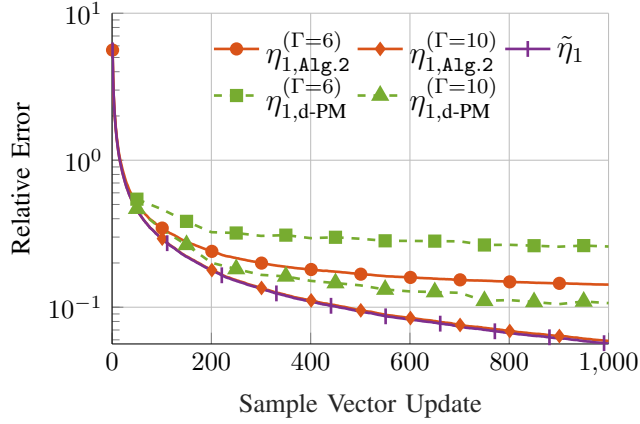}
	\caption{Relative error of $\lambda_1$ of $\mathbf{R}(t)$, where $\eta_{\mathtt{Alg. \ref{alg:onlineDis}}}$, $\eta_{\text{d-PM}}$, and $\eta_{\text{d-Oja}}$ stand for relative errors using Alg. \ref{alg:onlineDis}, using d-PM ($\Omega=30$) with AC protocol, and using d-NOja method ($\beta = 0.0005$), respectively. $\Gamma$ is the number of consensus iterations.} 
	\label{fig:covError}
\end{figure}

We first consider the conventional finite sample estimate for stationary signals which can be obtained by choosing $\alpha = \frac{t-1}{t}$ in (\ref{equ:sampleCov}). The d-PM is used as a comparison with the number of PM iterations $\Omega$. Another comparison is the decentralized Normalized Oja (d-NOja) method \cite{a100} with the step size $\beta$, which is a generalized decentralized Oja method for multiple principal eigenvectors. The relative errors are defined as
\begin{equation}
		\eta_k(t) = \frac{|\lambda_k(t) - \hat{\lambda}_k(t)|}{|\hat{\lambda}_k(t)|},\ \ \tilde{\eta}_k(t) = \frac{|\tilde{\lambda}_k(t) - \hat{\lambda}_k(t)|}{|\hat{\lambda}_k(t)|},
\end{equation}
where $\lambda_{k}(t)$ is the $k$-th eigenvalue computed by our proposed algorithm, $\tilde{\lambda}_k(t)$ is the $k$-th eigenvalue computed by a centralized processor, and $\hat{\lambda}_k$ is the $k$-th eigenvalue of the true sample distribution. Figure \ref{fig:covError} shows the relative error performance of the largest eigenvalue, i.e., $\lambda_1$, of $\mathbf{R}(t)$ with different numbers of consensus iterations $\Gamma$, where we observe that all distributed algorithms approach the relative error performance of the centralized algorithm.

We stress that although the relative error performance associated with the d-PM and d-NOja method is comparable to that of our distributed algorithm, the total numbers of consensus rounds\footnote{One consensus round is one communication instant that each node reaches the consensus of one scalar value with all its neighbors} of the d-PM and the d-NOja method are higher than that of our proposed scheme. More precisely, for an undirected network with $N$ nodes and $T$ sample vectors, our distributed algorithm requires $C_{\mathtt{Alg. \ref{alg:onlineDis}}} = 2NT$ consensus rounds for the estimation of all eigenvalues and eigenvectors, while the d-PM and the d-NOja method require higher numbers of total consensus rounds, i.e., $C_{\text{d-PM}}=N(T\Omega+T+2) + \Omega N(N-1)/2$ for the d-PM (including the associated distributed normalization and the largest eigenvalue subtraction), and $C_{\text{d-NOja}} = 2(N+N^2)T+2NT$ for the d-NOja method. Here, one PS round is treated as two AC rounds since two values are exchanged in each PS iteration. 

Furthermore, the d-PM is a batch algorithm that requires the knowledge of all sample vectors to perform the eigenvalue decomposition, whereas our distributed algorithm is an online algorithm that can update the eigenvalue each time when a new sample vector is obtained. Although the d-NOja method is an online approach for the eigenvector estimation, it still requires the knowledge of all sample vectors to compute the associated eigenvalue. Note that the PM and the Oja's method usually converge slowly in particular in the case where the largest eigenvalue is not dominant over the second largest eigenvalues \cite{a138,a131}, i.e., a larger eigenvalue gap or eigengap, which our distributed algorithm does not suffer from. The comparison between the proposed Algorithm \ref{alg:onlineDis} and the state-of-the-art d-PM and d-NOja method are summarized in Table \ref{tab:diff}.
\begin{table*}[t]
	\caption{Comparison between the proposed algorithm and the state-of-the-art d-PM and d-NOja method}
	\centering
	\begin{threeparttable}
		\begin{tblr}{c|c|c|c|c}
			Algorithm 		& Online Approach& Eigengap Assumption & Total Communication Cost	 & Local Storage\\
			\hline
			proposed Alg.\ref{alg:onlineDis}& yes			& no & $2NT$ 			 & $6N$	\\
			\hline
			d-PM     		&  no	 & yes		& $N(T\Omega+T+2) + \Omega N(N-1)/2$	 & $2N+T$\\
			\hline
			d-NOja & yes\tnote{*} & yes & $2(N+N^2)T + 2NT$	& $2N +T$\\
			\hline
		\end{tblr}	
		\begin{tablenotes}\centering
			\item[*] \vspace{2pt}\parbox{.75\linewidth}{\scriptsize For the computation of the eigenvalues, the d-NOja method requires the full knowledge of the sample vectors. Hence, it is an online approach for eigenvector estimation, and a batch approach for eigenvalue estimation.}
		\end{tablenotes}
	\end{threeparttable}
	
	\label{tab:diff}
\end{table*}

\subsection{Distributed Direction-of-Arrival (DoA) Estimation}\label{subsec:doa}
Computing the eigenvalues and eigenvectors of the sample covariance matrix distributively enables the distributed DoA estimation. Among various DoA estimation methods, the ESPRIT algorithm is particularly useful for distributed implementation, since the displacement of the subarrays can be arbitrary and unknown while only the inner subarray sensor locations must be known. Hence, we propose a distributed ESPRIT DoA estimation method based on our distributed online eigendecomposition algorithm, where the d-ESPRIT algorithm based on the d-PM \cite{a119} is mentioned as a comparison.

Considering a sensor network consisting of $N$ nodes with in total $M$ identically oriented antennas. The sensor network is a shift invariant system where all antennas can be categorized into the identical upper and lower group with the relative distance $\delta$. The shift invariant pairs from the upper group and the lower group form a subarray as a node in the network. Multiple pairs can merge together as a shift invariant subarray (node) consisting of multiple antennas.

The ESPRIT algorithm is a subspace based DoA estimation method and the eigenvalue decomposition of the sample covariance matrix $\mathbf{R}(t)$ can be partitioned as
\begin{equation}
	\begin{aligned}
		\mathbf{R}(t) 
		=\mathbf{U}_s(t)\boldsymbol{\Lambda}_s(t)\mathbf{U}_s(t)^\mathsf{H} + \mathbf{U}_n(t)\boldsymbol{\Lambda}_n(t)\mathbf{U}_n(t)^\mathsf{H},
	\end{aligned}
\end{equation}
where $\boldsymbol{\Lambda}_s(t)\in\mathbb{R}^{n\times n}$ and $\boldsymbol{\Lambda}_n(t)\in\mathbb{R}^{(M-n)\times (M-n)}$ are the diagonal matrices representing the signal and noise eigenvalues at the time instant $t$, respectively, $n$ is the number of DoAs, and $\mathbf{U}_s(t)\in\mathbb{C}^{M\times n}$ and $\mathbf{U}_n(t)\in\mathbb{C}^{M\times (M-n)}$ are the corresponding signal and noise eigenvectors, respectively. Based on the partition of the antennas in the subarrays, all the signal eigenvectors correspond to the upper group and the lower group are collected in $\overline{\mathbf{U}}_s(t)\in\mathbb{C}^{(M-N)\times n}$ and $\underline{\mathbf{U}}_s(t)\in\mathbb{C}^{(M-N)\times n}$, respectively. Following the conventional ESPRIT algorithm \cite{a3}, the DoAs are revealed by the eigenvalues of the matrix
\begin{equation}\label{equ:psi}
	\boldsymbol{\Psi} = \left(\overline{\mathbf{U}}_s(t)^\mathsf{H}\overline{\mathbf{U}}_s(t)\right)^{-1}\overline{\mathbf{U}}_s(t)^\mathsf{H}\underline{\mathbf{U}}_s(t)
\end{equation}
i.e., the DoAs are computed as
\begin{equation}
	\theta_i = \arcsin \left(-\frac{\arg(\psi_i)}{2\pi \delta}\right),\quad i = 1, \ldots, n,
\end{equation}
where $\psi_i$ are the eigenvalues of the matrix $\boldsymbol{\Psi}$.

As described in Section \ref{subsec:onlineAlg} and shown above in Section \ref{subsec:cov}, the eigenvectors of the sample covariance matrix are computed with our distributed online eigendecomposition algorithm, where each subarray has access to the corresponding rows of the matrices $\overline{\mathbf{U}}_s(t)$ and $\underline{\mathbf{U}}_s(t)$. %
Defining $\mathbf{C} = \overline{\mathbf{U}}_s(t)^\mathsf{H}\overline{\mathbf{U}}_s(t)$ and $\mathbf{F} = \overline{\mathbf{U}}_s(t)^\mathsf{H}\underline{\mathbf{U}}_s(t)$, equation \eqref{equ:psi} becomes
\begin{equation}
	\mathbf{C}\boldsymbol{\Psi} = \mathbf{F}.
\end{equation}
We observe that the computation of any entry in the matrix $\mathbf{C}$ is expressed explicitly as 
\begin{equation}
	\begin{aligned}
		c_{jk} &= \bar{\mathbf{u}}_{s,j}(t)^\mathsf{H}\bar{\mathbf{u}}_{s,k}(t)\\
		 &= \sum_{i = 1}^{N}\bar{\mathbf{u}}_{s,\{j,i\}}(t)^\mathsf{H}\bar{\mathbf{u}}_{s,\{k,i\}}(t)\quad \forall j,k\in \{1,2,\ldots,n\},
	\end{aligned}
\end{equation}
where $\bar{\mathbf{u}}_{s,j}(t)$ and $\bar{\mathbf{u}}_{s,k}(t)$ are the $j$-th and $k$-th column of the matrix $\overline{\mathbf{U}}_s(t)$, respectively, and $\bar{\mathbf{u}}_{s,\{j,i\}}(t)$ and $\bar{\mathbf{u}}_{s,\{k,i\}}(t)$ are the corresponding components associated with the $i$-th node, which are known locally. Therefore, the entry $c_{jk}$ of the matrix $\mathbf{C}$ can be computed distributively with any consensus protocols introduced in Section \ref{sec:protocols}. Since the size of the matrix $\mathbf{C}$ is $n\times n$, only $n^2$ consensus rounds are required to compute the matrix $\mathbf{C}$. A similar approach can be applied to the computation of the entries of the matrix $\mathbf{F}$ with additional $n^2$ consensus rounds. Consequently, the matrices $\mathbf{C}$ and $\mathbf{F}$ will be available at each subarray. Hence, the eigenvalues of the matrix $\boldsymbol{\Psi}$ are computed locally at each subarray, and the DoAs are found. Note that the communication and computation cost related to $\boldsymbol{\Psi}$ are low since the matrices $\mathbf{C}$, $\mathbf{F}$ and $\boldsymbol{\Psi}$ are of size $n\times n$, which are only related to the number of sources. 

To illustrate and compare the performance of the distributed DoA estimation using our distributed online eigenvalue decomposition algorithm with the d-ESPRIT algorithm, we adopt the same simulation setup as in \cite{a119}, with $N=6$ subarrays, each of which consists of two antennas separated by half a wavelength, i.e., $M_k = 2$, for $k = 1, \ldots, 6$. Thus, the upper group and the lower group consist of the first antenna (i.e., the reference antenna) and the second antenna in each subarray, respectively. The subarrays are connected in a way that the neighboring sets are $\mathcal{N}_1 = \{2, 3\}$, $\mathcal{N}_2 = \{1, 3\}$, $\mathcal{N}_3 = \{1, 2, 4\}$, $\mathcal{N}_4 = \{3, 5, 6\}$, $\mathcal{N}_5 = \{4, 6\}$, and $\mathcal{N}_6 = \{4, 5\}$ as illustrated in Figure \ref{fig:subarrays}.
In total $200$ snapshots from $n = 3$ sources nearby located at $-7, 19$ and $23$ degrees are available for the DoA estimation. The Root Mean Square Error (RMSE) performance over $100$ Monte Carlo iterations is shown in Figure \ref{fig:rmse}.
\begin{figure}[t]
	\begin{minipage}{.45\linewidth}
		\centering
		\begin{tikzpicture}		
	\definecolor{mycolor1}{RGB}{255,0,0} 
	
	\definecolor{mycolor2}{RGB}{0,255,0} 
	\definecolor{mycolor3}{RGB}{0,0,255} 
	\definecolor{mycolor4}{RGB}{0,0,0}   
	\definecolor{mycolor5}{RGB}{0, 128, 128} 
	\definecolor{mycolor6}{RGB}{90, 12, 90} 
	\definecolor{mycolor7}{RGB}{0,255,255} 
	\definecolor{mycolor8}{RGB}{255,255,0} 
	\definecolor{mycolor9}{RGB}{255,0,255} 
	\definecolor{mycolor10}{RGB}{142, 37, 37} 
	\definecolor{mycolor11}{RGB}{255, 165, 0} 
	\definecolor{mycolor12}{RGB}{255, 51, 204} 
	\definecolor{mycolor13}{RGB}{128, 128, 0} 
			

	%
	\def\ant#1{ 
		\node at ($ (#1.north) + (-.3,-0.12)$)(a) {};
		\draw[fill=black!85!white] (a)  -- ++(0,0.32) --   ++(-.14,.2) -- ++(.3,0) -- ++(-.14,-.2) -- ++(0,-.2) -- ++(0.6,0) -- ++(0,0.2) -- ++(-.14,.2) -- ++(.3,0) -- ++(-.14,-.2) -- ++(0,-.2);
	};

	\def\agent#1#2#3{
		\node[rectangle,draw,fill=black!15!white](#1) at #2 {\scriptsize{Subarray #3}};
		\ant{#1};
	}

	\foreach \i/\x in {{6/(2.9,-1.1)},{5/(4.2,.4)},{4/(1.8,.2)},{3/(.7,-1.4)}, {2/(-.6,1.1)},{1/(-1.9,-.5)}}
	\agent{\i}{\x}{\i};

	\begin{scope}[on background layer]
		
		\draw ([xshift=14]1.north) -- (2);
		\draw (1) -- (3.west);
		\draw (2) -- ([xshift=-15]3.north);
		\draw ([xshift=15]3.north) -- (4);
		\draw (4) -- (5);
		\draw (4) -- ([xshift=-10]6.north);
		\draw (5) -- ([xshift=15]6.north);
	\end{scope}

\end{tikzpicture}	
		\caption{The sensor network for DoA estimation with $N = 6$ subarrays, where each subarray contains $M_k = 2$ antennas.}
		\label{fig:subarrays}
	\end{minipage}
	\hfil
	\begin{minipage}{.45\linewidth}
		\centering
		\definecolor{mycolor1}{rgb}{0.00000,0.44700,0.74100}%
\definecolor{mycolor2}{rgb}{0.85000,0.32500,0.09800}%
\definecolor{mycolor3}{rgb}{0.92900,0.69400,0.12500}%
\definecolor{mycolor4}{rgb}{0.49400,0.18400,0.55600}%
\definecolor{mycolor5}{rgb}{0.46600,0.67400,0.18800}%
\definecolor{mycolor6}{rgb}{0.30100,0.74500,0.93300}%
\begin{tikzpicture}
	
	\begin{axis}[
		width=6.6cm,
		height=4.4cm,
		at={(0,0)},
		scale only axis,
		xmin=-10,
		xmax=50,
		ymode=log,
		ymin=1e-2,
		ymax=1e2,
		legend style={legend cell align=left, align=left, draw=none, fill=none, nodes={scale=0.85}, font=\footnotesize},
		legend columns = 2,
		xmajorgrids,
		ymajorgrids,
		xlabel style={font=\color{white!15!black}},
		xlabel=SNR (dB),
		ylabel style={font=\color{white!15!black}},
		ylabel=RMSE (deg),
		]
		
\pgfplotstableread{
	x		y	
	-10  1.8284
	-5  1.59077
	0  0.953421
	5  0.526339
	10  0.212241
	15  -0.0206793
	20  -0.281498
	25  -0.521867
	30  -0.775726
	35  -1.03279
	40  -1.28204
	45  -1.51441
	50  -1.71115
}{\mytabled};
\addplot[line width = 1pt, smooth,mark=*, mycolor2] table[row sep=crcr, y expr=10^\thisrow{y}] {\mytabled};					
\addlegendentry{$\text{Alg. \ref{alg:onlineDis}}(\Gamma = 10)$}

\pgfplotstableread{
	x		y	
	-10  1.75272
	-5  1.36885
	0  0.800648
	5  0.456366
	10  0.172311
	15  -0.0649146
	20  -0.320073
	25  -0.564315
	30  -0.817585
	35  -1.0774
	40  -1.32556
	45  -1.55611
	50  -1.76371
}{\mytablee};

\addplot[line width = 1pt, smooth,mark=diamond*, mycolor2] table[row sep=crcr, y expr=10^\thisrow{y}] {\mytablee};
\addlegendentry{$\text{Alg. \ref{alg:onlineDis}}(\Gamma = 15)$}

\pgfplotstableread{
	x		y	
	-10  1.68912
	-5  1.42398
	0  0.90775
	5  0.447111
	10  0.288696
	15  0.106735
	20  0.00715865
	25  -0.100727
	30  -0.264991
	35  -0.37419
	40  -0.41742
	45  -0.40809
	50  -0.424778
}{\mytableb};
\addplot[line width = 1pt, smooth,mark=square*, mycolor5] table[row sep=crcr, y expr=10^\thisrow{y}] {\mytableb};
\addlegendentry{$\text{d-PM}(\Gamma = 10)$}

\pgfplotstableread{
	x		y	
	-10  1.63769
	-5  1.33096
	0  0.863858
	5  0.47372
	10  0.174786
	15  0.0950647
	20  -0.173796
	25  -0.349547
	30  -0.475942
	35  -0.574497
	40  -0.699148
	45  -0.792907
	50  -0.815968
}{\mytablec};
\addplot[line width = 1pt, smooth,mark=triangle*,mycolor5] table[row sep=crcr, y expr=10^\thisrow{y}] {\mytablec};				
\addlegendentry{$\text{d-PM}(\Gamma = 15)$}


\pgfplotstableread{
	x		y	
	-10  1.49592
	-5  1.14551
	0  0.733518
	5  0.400123
	10  0.126989
	15  -0.120534
	20  -0.367441
	25  -0.600409
	30  -0.868433
	35  -1.11293
	40  -1.36337
	45  -1.61357
	50  -1.86366
}{\mytablea};

\addplot[line width = 1pt, smooth,mark=|, black] table[row sep=crcr, y expr=10^\thisrow{y}] {\mytablea};
\addlegendentry{$\text{ESPRIT}$}

\pgfplotstableread{
	x		y	
	-10 1.41447
	-5  0.978712
	0   0.598221
	5   0.280186
	10  0.00289705
	15  -0.256561
	20  -0.509647
	25  -0.760633
	30  -1.01095
	35  -1.26105
	40  -1.51107
	45  -1.76108
	50  -2.01108
}{\mytablea};

\addplot[line width = 1pt, smooth,mark=, purple] table[row sep=crcr, y expr=10^\thisrow{y}] {\mytablea};
\addlegendentry{$\text{CRB}$\cite{a126}}
		
	\end{axis}
\end{tikzpicture}
		\caption{RMSE performance of d-ESPRIT using Alg. \ref{alg:onlineDis} and d-PM ($\Omega = 4$), with the number of PS iterations in both algorithms $\Gamma = 10$ and $\Gamma = 15$. The RMSE is computed over $200$ snapshots and $100$ Monte Carlo iterations.}
		\label{fig:rmse}
	\end{minipage}
\end{figure}

We observe that both distributed algorithms achieve the RMSE performance of the centralized ESPRIT algorithm. However, for high SNR scenarios, i.e., $\text{SNR} \geq 20 \text{dB}$, the d-ESPRIT algorithm using our distributed algorithm still achieves the RMSE performance of the centralized ESPRIT algorithm, whereas the d-ESPRIT algorithm using d-PM has a larger RMSE both for consensus iterations $\Gamma = 10$ and $\Gamma = 15$.
\begin{figure}[t]
	\begin{minipage}{.45\linewidth}
		\centering
		\vspace{1.6cm}
		\input{simulations/DOA_tracking1}
		\caption{DoA tracking with two moving sources. $\text{SNR} = 20 \ \text{dB}$, the number of PS iterations $\Gamma = 15$ and $\alpha = 0.88$. $\text{RMSE} = 1.111$.}
		\label{fig:doatracking}
	\end{minipage}
	\hfil
	\begin{minipage}{.45\linewidth}
		\centering
		\vspace{1.7cm}
		\input{simulations/DOA_trackingOja}
		\caption{DoA tracking with two moving sources using the d-NOja method. $\text{SNR} = 20 \ \text{dB}$, the number of PS iterations $\Gamma = 15$ and $\beta = 0.02$. $\text{RMSE} = 2.2312$.}
		\label{fig:doatrackingOja}
	\end{minipage}
\end{figure}

Apart from estimating the DoAs of stationary sources, our proposed d-ESPRIT algorithm using the distributed online eigenvalue decomposition approach naturally enables us to track the moving sources by simply updating the newly obtained sample vectors, e.g., by setting the forgetting factor $\alpha$ in (\ref{equ:sampleCov}) as a constant. In Figure \ref{fig:doatracking}, two sources move on crossing trajectories. The forgetting factor is chosen as $\alpha = 0.88$ and we observe that our proposed DoA tracking scheme successfully estimates and tracks the DoAs of the two moving sources.

As a comparison, we also implement the d-NOja method \cite{a100}. While the d-NOja method successfully estimates and tracks the DoAs as shown in Figure \ref{fig:doatrackingOja}, however, it suffers from slow convergence speed, and thus, higher RMSE over all samples compared to the DoA tracking performance of our algorithm as illustrated in Figure \ref{fig:doatracking}.

Furthermore, due to the fact that the d-PM is a batch algorithm that requires all the sample vectors at once to estimate the sample covariance matrix, the algorithm does not naturally extend to an online tracking implementation for moving targets. One possible method to perform the DoA tracking with the d-PM is by applying a sliding window as indicated in (\ref{equ:sampleCov2}), where an old sample outside the sliding window is dropped while a new one is obtained. However, in this way, the d-PM still suffers from high communication cost to perform the DoA estimation in each sliding window. Nevertheless, the sliding window approach can be carried out according to \eqref{equ:sampleCov2} as two consecutive rank-one modifications. Hence, our distributed algorithm is still applicable with significantly lower communication costs compared to the d-PM. As a matter of fact, only the two rank-one updates need to be communicated over the network with $C_{\mathtt{Alg. \ref{alg:onlineDis}},\mathtt{window}} = 2N$ consensus rounds in each sliding window update by performing our proposed distributed algorithm, whereas the d-PM requires a whole update of all the sample vectors with $C_{\text{d-PM},\mathtt{window}} = N(\beta\Omega+\beta+2) + \Omega N(N-1)/2$ consensus rounds for a window length $\beta$.

\subsection{Distributed Graph Spectrum Estimation and Tracking}\label{subsec:lap}
We consider now the application example of the distributed computation of the graph spectrum, hence, the eigenvalues of the graph Laplacian $\mathbf{L}$. The graph Laplacian is often used as the shift operator in GSP, and its eigenvalues not only reveal the characteristic of the corresponding graph but can also be helpful in designing graph filters and filter banks, etc \cite{a134}. However, the centralized eigendecomposition of $\mathbf{L}$ requires a large communication and coordination overhead, particularly in large scale and evolving networks. To reduce this overhead and to make the tracking of the graph spectrum scalable we propose a distributed spectrum computation approach.

\subsubsection{Spectrum Estimation in Static Networks}
In this application example, we want to compute the eigenvalues of the graph Laplacian of a static network, where each node is labeled and interacts only with its neighboring nodes. We assume that the network is synchronized so that the communication is accomplished within the same synchronized time slot. To compute the eigenvalues of the graph Laplacian by our proposed algorithm, we express $\mathbf{L}$ as
\begin{equation}\label{eq_L_update}
	\mathbf{L} = \mathbf{B}\mathbf{B}^\mathsf{T} = \sum_{g=1}^{N_e}\mathbf{b}_g\mathbf{b}_g^\mathsf{T},\myvspace
\end{equation}
where $\mathbf{B}=[\mathbf{b}_1,\ldots,\mathbf{b}_{N_e}]\in\mathbb{R}^{N\times N_e}$ is the oriented incidence matrix with $N_e = |\mathcal{E}|$. 
The entry $b_{kg}$ for the $k$-th node and the $g$-th edge (connecting the $i$-th node and the $j$-th node) is\myvspace
\begin{equation}
	b_{kg} = \begin{cases}
		1, \quad &\text{if}\ k = i,\ g = j,\\
		-1, \quad &\text{if}\ k = j,\ g = i,\\
		0, \quad &\text{otherwise}.
	\end{cases}\myvspace
\end{equation}  

In order to enable the nodes to cooperatively update the graph Laplacian with rank-one modifications according to \eqref{eq_L_update}, an appropriate protocol is required, where all edges must be visited exactly once and the $k$-th node, for $k\in\mathcal{V}$, sends its value $b_{kg}\in\{1, -1, 0\}$ of the incidence matrix $\mathbf{B}$, for $g = 1, \ldots, N_e$, as the graph signal correspondingly. Since each node has the access to all its connected edges and can exchange information with its adjacent neighbors, a protocol can be designed by finding the minimum spanning tree or any spanning tree of the graph, where a path through all nodes can be found. The distributed minimum spanning tree problem is well studied and various distributed algorithms are available in the literature, see \cite{a114,a115,a116,a117} and the references therein.%

As a matter of fact, only a subset of the minimum spanning tree already satisfies our need as long as all edges are covered. Since not all nodes are necessary to be visited to guarantee the coverage of all edges, we provide here a simple protocol to find a sequence of nodes from which all edges in the graph can be reached.

\paragraph*{Node Sequence Determination Protocol}
In the network, a node can be the \textit{Head}, which sends $1$, or the \textit{Tail}, which sends $-1$, of an edge. There is only one active \textit{Head} and one active \textit{Tail} in the network in each synchronized time slot. If a node is the active \textit{Head}, it cooperates with its neighbors to update the graph Laplacian with each of its edges in the network, and then the process is repeated in one of its adjacent neighbors. A node should not be the active \textit{Tail} if it has already been the active \textit{Head}. When there are no available neighbors, i.e., all the neighbors have been the active \textit{Head}, the process is handed over to the previous active \textit{Head} until it reaches the first active \textit{Head}, and thus the protocol terminates. Different from the (minimum) spanning tree problem, in this protocol, not all nodes have to be visited, thus, to be the active \textit{Head}, but all edges have been examined once.$\hfill \blacksquare$
\begin{figure}[t]
	\centering
%
%
\definecolor{mycolor1}{rgb}{0.00000,0.44700,0.74100}%
\definecolor{mycolor2}{rgb}{0.85000,0.32500,0.09800}%
\begin{tikzpicture}

\begin{axis}[%
width=6.6cm,
height=4.4cm,
at={(0,0)},
scale only axis,
xmin=50,
xmax=300,
xlabel style={font=\color{white!15!black}},
xlabel={Network Size $N$},
xtick={50,100,150,200,250,300},
ymode=log,
ymin=1e-06,
ymax=1,
ytick={10e-6,10e-4,10e-2},
yminorticks = true,
axis x line*=bottom,
axis y line*=right,
ylabel={Relative Error (solid)},
xmajorgrids,
ymajorgrids
]
\addplot [line width = 1pt,color=mycolor1,mark = triangle*,mark size = 2pt]
  table[row sep=crcr]{%
50	1.22224860970797e-06\\
100	1.6524342329911e-06\\
150	1.27488780627481e-05\\
200	1.10477315530149e-05\\
250	1.9580017066678e-05\\
300	4.12441975454799e-05\\
};

\addplot [line width = 1pt,color=mycolor2,mark = diamond*,mark size = 2pt]
  table[row sep=crcr]{%
50	0.0159970303684138\\
100	0.0217591081685059\\
150	0.0236478816017009\\
200	0.0261411665761005\\
250	0.024788500624985\\
300	0.0271032194454094\\
};

\end{axis}
\begin{axis}[%
width=6.6cm,
height=4.4cm,
at={(0,0)},
scale only axis,
xmin=50,
xmax=300,
xlabel style={font=\color{white!15!black}},
xtick={50,100,150,200,250,300},
ymin=0,
ymax=6000000,
axis x line*=bottom,
axis y line*=left,
ylabel={Communication Cost (dashed)},
xmajorgrids,
	ymajorgrids,
legend columns = 2,
legend style={fill opacity=1, legend pos=north west,legend cell align=left, align=left, draw=none, fill=none}
]
\addlegendimage{/pgfplots/refstyle=plot_one,color=mycolor1,line width = 1pt,mark = triangle*,mark size = 2pt}\addlegendentry{\large$\eta_{1,\mathtt{Alg. \ref{alg:onlineDis}}}$}
\addlegendimage{/pgfplots/refstyle=plot_two,color=mycolor2,line width = 1pt,mark = diamond*,mark size = 2pt}\addlegendentry{\large$\eta_{1,\text{d-PM}}$}
\addplot [line width = 1pt,color=mycolor1,dashed,mark=square*,mark size = 2pt,mark options={fill=mycolor1,solid}]
  table[row sep=crcr]{%
50	60000\\
100	120000\\
150	180000\\
200	240000\\
250	300000\\
300	360000\\
};
\addlegendentry{\large$C_{\mathtt{Alg. \ref{alg:onlineDis}}}$}

\addplot [line width = 1pt,color=mycolor2, dashed,mark=*,mark size = 2pt,mark options={fill=mycolor2,solid}]
  table[row sep=crcr]{%
50	654600\\
100	1359200\\
150	2113800\\
200	2918400\\
250	3773000\\
300	4677600\\
};
\addlegendentry{\large$C_{\text{d-PM}}$}

\end{axis}

\end{tikzpicture}%
	\caption{Relative error of $\lambda_1$ and total communication cost of all eigenvalues for graph Laplacian of $d$-regular ($d = 4$) networks with different network size $N$, where the number of PS iterations $\Gamma = 100$ and $\Omega = 20$.}
	\label{fig:laplacianError}
\end{figure}
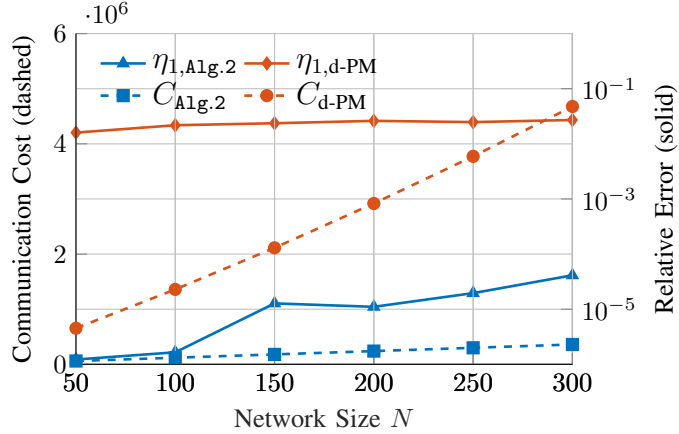

Along the node sequence found by the aforementioned protocol, the eigendecomposition of $\mathbf{L}$ can be carried out using Algorithm \ref{alg:onlineDis}. %
Apart from the relative error performance, we also examine the communication cost of our distributed algorithm with different network sizes, and compare both with the d-PM, which are shown in Figure \ref{fig:laplacianError}. We observe that the communication cost of the d-PM increases dramatically when the network size increases and the relative error performance remains similar, whereas our distributed algorithm reaches a better relative error performance with a reasonable communication cost increase.

\subsubsection{Dynamic Graph Spectrum Tracking}
Another advantage of our distributed algorithm compared to the d-PM is the ability to efficiently track eigenvalues of graph Laplacian matrices associated with dynamically evolving graphs. Denote a given graph Laplacian $\mathbf{L}(t-1)$ at the time instant $(t-1)$. Further assume that at the time instant $t$, a random edge disappears in the network, though the network remained to be connected. Then we can use equation \eqref{equ:problem} with $\mathbf{R}(t-1) = \mathbf{L}(t-1)$, $\mathbf{x}(t) = \mathbf{b}_\ell$ and $\rho(t) = -1$ to express the graph Laplacian at the time instant $t$ as a rank-one update. Similarly, if at the time instant $t$ a new edge appears in the network we choose $\rho(t) = 1$ in the update.

\paragraph*{Strategy Dealing With Node Variation}
Apart from the edge variation, a node can appear or disappear in an evolving network. If a node is added to the existing network, it triggers the update process by adding its associated edges one by one. Similarly, if a node leaves the network in a controlled manner (e.g., it informs the network before it leaves), it removes its edges one by one. {Nevertheless, a node may leave the network without notification, e.g., due to node failure. In such cases, we assume that each node maintains a one-hop-neighbor table, i.e., each node has the knowledge of the neighbor sets of its neighbors. Once a node fails, its neighboring nodes carry out the update process according to the table to remove the associated edges.}$\hfill\blacksquare$
\begin{figure}[t]
	\begin{minipage}{.41\linewidth}
		\centering
		\input{simulations/case2_2}
		\caption{Eigenvalue learning and adaptation for dynamic graph network with $N = 50$ nodes and the number of PS iterations $\Gamma = 100$.}
		\label{fig:laplacianUpdate}
	\end{minipage}
	\hfil
	\begin{minipage}{.41\linewidth}
		\centering
		\input{simulations/case3}
		\caption{Eigenvalue learning and adaptation for dynamic graph network with $N = 50$ nodes and the number of PS iterations $\Gamma = 100$. The learning phase is reduced to $N$.}
		\label{fig:laplacianUpdate2}
	\end{minipage}
\end{figure}

Figure \ref{fig:laplacianUpdate} shows the largest eigenvalue tracking when the network evolves (edge appearing and disappearing), where the network is initialized as an undirected $d$-regular network with $d = 4$ and $N = 50$ nodes. For the first $100$ iterations, our distributed algorithm evaluates over all existing edges and computes the eigenvalues of the graph Laplacian. Then, at iteration $101$ the network starts evolving, i.e., edges are randomly removed or added, and the relative error behavior shows that our distributed algorithm is able to track the evolution of the network in the eigenvalues.%

\subsubsection{Communication Efficient Spectrum Estimation}
With the increase of the size or the density of the network, the number of edges increases, thus the number of iterations that are needed for our distributed algorithm to learn the network topology increases since all edges must be taken into account to compute the eigenvalues of the graph Laplacian.

One way to reduce the required number of iterations in the learning phase shown above is directly examining the graph Laplacian $\mathbf{L}$ instead of the incidence matrix $\mathbf{B}$. Thus, we introduce a two-step consecutive rank-one update procedure to recursively estimate $\mathbf{L}$ as
\begin{equation}
	\mathbf{L}(t) = \mathbf{L}(t-1) + \tilde{\mathbf{x}}(t)\tilde{\mathbf{x}}(t)^\mathsf{T} -\bar{\mathbf{x}}(t)\bar{\mathbf{x}}(t)^\mathsf{T}, \ t = 1,\ldots,N,
\end{equation}
where $\mathbf{L}(0)=\mathbf{0}_{N\times N}$,
\begin{equation}\label{equ:rank2-1}
	\tilde{\mathbf{x}}(t) = \begin{bmatrix}
		\mathbf{0}_{t-1}\\
		\sqrt{\ell_{t,t}}\\
		\frac{\ell_{t+1,t}}{\sqrt{\ell_{t,t}}}\\
		\vdots\\
		\frac{\ell_{N,t}}{\sqrt{\ell_{t,t}}}
	\end{bmatrix}\quad \text{and}\quad
	\bar{\mathbf{x}}(t) = \begin{bmatrix}
		\mathbf{0}_{t}\\
		\frac{\ell_{t+1,t}}{\sqrt{\ell_{t,t}}}\\
		\vdots\\
		\frac{\ell_{N,t}}{\sqrt{\ell_{t,t}}}	\end{bmatrix}.
\end{equation}
Note that in this procedure the time instant index $t$ also corresponds to the edge index. It is easy to show that $\mathbf{L}(N) = \mathbf{L}$.
In such a way, the learning phase of the spectrum computation of a graph is restricted to a total of $N$ iterations as compared to $N_e$ iterations in the case of the update based on the incidence matrix. When the number of edges $N_e$ exceeds the number of nodes, the scheme based on $\mathbf{L}$ is favorable to the update based on $\mathbf{B}$. Moreover, the $j$-th element in $\tilde{\mathbf{x}}(t)$ and $\bar{\mathbf{x}}(t)$ is either directly accessible at the corresponding node ($\ell_{j,t}$) or can be obtained from its adjacent neighbors ($\ell_{t,t}$ for $j\neq t$). As mentioned in Section \ref{subsec:doa}, such consecutive rank-one updates can also be solved by our distributed algorithm.

The track of the largest eigenvalue of the graph Laplacian of an evolving network is shown in Figure \ref{fig:laplacianUpdate2}, where the iteration of the learning phase is reduced to the number of nodes $N$ instead of the number of edges $N_e$ as shown in Figure \ref{fig:laplacianUpdate}. 

\textit{Remark:} The application of spectrum computation facilitated with the rank-one modification is not only suitable to the graph Laplacian itself but also applicable for its variations, such as the symmetric normalized graph Laplacian $\widetilde{\mathbf{L}}$ \cite{a135}, which can be expressed as $	\widetilde{\mathbf{L}} = \mathbf{I} - \mathbf{D}^{-\frac{1}{2}}\mathbf{A}\mathbf{D}^{-\frac{1}{2}} = \widetilde{\mathbf{B}}\widetilde{\mathbf{B}}^\mathsf{T}$ 
with $\widetilde{\mathbf{B}} = \mathbf{D}^{-\frac{1}{2}}\mathbf{B}$.

\subsection{Distributed Eigenvalue Decomposition of the Sample Covariance Matrix with Adapted Graph Laplacian.}\label{subsec:cov2}
As can be observed in Section \ref{subsec:cov}, \ref{subsec:doa} and \ref{subsec:lap}, the application examples using our distributed scheme shown in Algorithm \ref{alg:onlineDis} require a large number of PS or AC iterations in the large network scenarios, and in theory, an infinite number of consensus iterations is needed to achieve exact convergence. According to Section \ref{sec:protocols}, one way to avoid this is by applying the ftAC protocol which reaches exact convergence in finite time. However, the ftAC requires the knowledge of the eigenvalues of the graph Laplacian. Nevertheless, as mentioned in Section \ref{subsec:lap}, we can estimate the eigenvalues of the graph Laplacian, and thus adapt the step size in the ftAC protocol for, e.g., the eigenvalue decomposition of the sample covariance matrix. Although this brings extra computation and communication overhead at the initialization step or when the structure of the graph changes, we can benefit from faster convergence in each rank-one update in the main application.

Similar to the application example shown in Section \ref{subsec:cov}, we evaluate the relative error achieved by our Algorithm \ref{alg:onlineDis} with the ftAC protocol using the adapted eigenvalues of the graph Laplacian. We choose a small network example with $N = 10$ nodes shown in Figure \ref{fig:network}, whose graph Laplacian has $5$ distinct non-zero eigenvalues.
\begin{figure}[t]
	\begin{minipage}{.45\linewidth}
		\centering
			\begin{tikzpicture}
		
		\definecolor{mycolor1}{RGB}{255,0,0} 
		
		\definecolor{mycolor2}{RGB}{0,255,0} 
		\definecolor{mycolor3}{RGB}{0,0,255} 
		\definecolor{mycolor4}{RGB}{0,0,0}   
		\definecolor{mycolor5}{RGB}{0, 128, 128} 
		\definecolor{mycolor6}{RGB}{90, 12, 90} 
		\definecolor{mycolor7}{RGB}{0,255,255} 
		\definecolor{mycolor8}{RGB}{255,255,0} 
		\definecolor{mycolor9}{RGB}{255,0,255} 
		\definecolor{mycolor10}{RGB}{142, 37, 37} 
		\definecolor{mycolor11}{RGB}{255, 165, 0} 
		\definecolor{mycolor12}{RGB}{255, 51, 204} 
		\definecolor{mycolor13}{RGB}{128, 128, 0} 

		
		\def\agent#1#2#3{
			\node[circle,draw,fill=yellow!15!white,inner sep=2,,minimum size=1pt](#1) at #2 {\scriptsize #3};
		}

		
		
		
		
		
		\foreach \i/\x in {{1/(0,0.9)},{2/(1.4,-.1)},{3/(.8,-1.3)},{4/(-.8,-1.3)}, {5/(-1.4,-0.1)},{6/(0,1.7)},{7/(2.1,.4)},{8/(1.4,-1.9)},{9/(-1.4,-1.9)}, {10/(-2.1,0.4)}}
		\agent{\i}{\x}{\i};

		\begin{scope}[on background layer]
			
			\draw (1) -- (2);
			\draw (1) -- (5);
			\draw (1) -- (6);
			\draw (2) -- (3);
			\draw (2) -- (7);
			\draw (3) -- (4);
			\draw (3) -- (8);
			\draw (4) -- (5);
			\draw (4) -- (9);
			\draw (5) -- (10);
		\end{scope}


	\end{tikzpicture}	
		\caption{Sample network with $N = 10$ nodes. Its graph Laplacian has $5$ distinct non-zero eigenvalues.}
		\label{fig:network}
	\end{minipage}
\hfill
	\begin{minipage}{.45\linewidth}
		\centering
		\input{simulations/case4}
		\caption{Relative error of $\lambda_{1}$ of $\mathbf{R}(t)$ using Alg. \ref{alg:onlineDis} with the AC algorithm and the PS algorithm, and the ftAC protocol with adapted step size based on the distributively computed eigenvalues of the graph Laplacian.}
		\label{fig:case4}
	\end{minipage}
\end{figure}

The relative error performance is shown in Figure \ref{fig:case4}, where the number of iterations required in the ftAC protocol is equal to the number of distinct nonzero eigenvalues which is $\Gamma = 5$ in this example. Figure \ref{fig:case4} shows that the achieved relative error performance based on the ftAC protocol with the adapted eigenvalues of the graph Laplacian surpasses those based on the AC and the PS protocols with the same number of iterations ($\Gamma = 5$) and even with twice the number of iterations ($\Gamma = 10$), and approaches the relative error performance computed with a central processor. This is attractive not only for the distributed eigenvalue decomposition of the sample covariance matrix but also for various decentralized algorithms that are based on the decentralized averaging protocols.

\subsection{Eigenvalue Dependent Graph Filter Design with Adapted Graph Laplacian.}\label{subsec:gfilter} 
Another way to avoid the infinite consensus iterations associated with the PS and the AC protocols is by applying the low-pass graph filter approach introduced in Section \ref{subsec:gfilterConsensus}. Similar to the ftAC protocol, the knowledge of the eigenvalues of the graph Laplacian is also required in the graph dependent filter design. Thus, apart from adopting the estimated eigenvalues of the graph Laplacian in the ftAC protocol, the eigenvalues can also be utilized in the graph dependent filter design. Taking advantage of the adapted graph spectrum introduced in Section \ref{subsec:lap}, we are able to design graph dependent filters based on the estimated graph frequencies as introduced in Section \ref{subsec:gfilterConsensus}, which also avoids the high computation cost of the direct eigendecomposition of the graph Laplacian.
We illustrate this by designing a low-pass finite impulse response (FIR) graph filter based on the learned eigenvalues of the graph Laplacian using the algorithms introduced in Section \ref{subsec:lap}. 

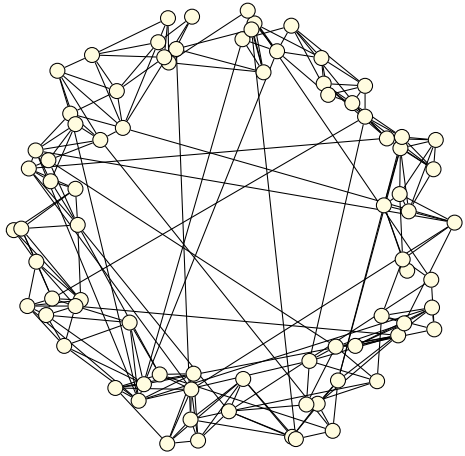
\begin{figure}[t]
	\centering
%
%
\begin{tikzpicture}
		\definecolor{mycolor1}{RGB}{255,0,0} 

\definecolor{mycolor2}{RGB}{0,255,0} 
\definecolor{mycolor3}{RGB}{0,0,255} 
\definecolor{mycolor4}{RGB}{0,0,0}   
\definecolor{mycolor5}{RGB}{0, 128, 128} 
\definecolor{mycolor6}{RGB}{90, 12, 90} 
\definecolor{mycolor7}{RGB}{0,255,255} 
\definecolor{mycolor8}{RGB}{255,255,0} 
\definecolor{mycolor9}{RGB}{255,0,255} 
\definecolor{mycolor10}{RGB}{142, 37, 37} 
\definecolor{mycolor11}{RGB}{255, 165, 0} 
\definecolor{mycolor12}{RGB}{255, 51, 204} 
\definecolor{mycolor13}{RGB}{128, 128, 0} 


\def\agent#1#2{
	\node[circle,draw,fill=yellow!15!white,inner sep=2,minimum size=1pt](#1) at #2 {};
}

\foreach \i/\x in {
	{1/(3.48,0.55)},
	{2/(2.54,0.78)},
	{3/(2.75,0.93)},
	{4/(2.87,0.70)},
	{5/(2.76,1.54)},
	{6/(3.23,1.65)},
	{7/(3.20,1.26)},
	{8/(2.78,1.69)},
	{9/(2.58,1.67)},
	{10/(2.29,1.96)},
	{11/(2.12,2.14)},
	{12/(1.74,2.40)},
	{13/(2.29,2.37)},
	{14/(1.80,2.25)},
	{15/(1.71,2.74)},
	{16/(1.31,3.17)},
	{17/(1.12,2.83)},
	{18/(0.82,3.20)},
	{19/(0.66,2.99)},
	{20/(0.73,3.37)},
	{21/(0.94,2.55)},
	{22/(0.78,3.12)},
	{23/(-0.32,2.68)},
	{24/(-0.22,2.86)},
	{25/(-0.01,3.29)},
	{26/(-0.33,3.27)},
	{27/(-0.38,2.75)},
	{28/(-0.46,2.95)},
	{29/(-1.34,2.78)},
	{30/(-1.01,2.30)},
	{31/(-0.93,1.81)},
	{32/(-1.80,2.57)},
	{33/(-1.63,2.00)},
	{34/(-1.23,1.65)},
	{35/(-1.56,1.86)},
	{36/(-2.09,1.51)},
	{37/(-1.92,1.38)},
	{38/(-1.89,1.10)},
	{39/(-2.18,1.27)},
	{40/(-1.56,1.00)},
	{41/(-1.53,0.52)},
	{42/(-2.38,0.45)},
	{43/(-2.28,0.47)},
	{44/(-1.49,-0.48)},
	{45/(-2.08,0.03)},
	{46/(-1.87,-0.46)},
	{47/(-2.20,-0.56)},
	{48/(-1.56,-0.56)},
	{49/(-1.95,-0.68)},
	{50/(-1.71,-1.09)},
	{51/(-0.84,-0.78)},
	{52/(-0.65,-1.60)},
	{53/(-0.72,-1.82)},
	{54/(-0.44,-1.47)},
	{55/(-1.03,-1.65)},
	{56/(0.01,-1.46)},
	{57/(-0.02,-1.67)},
	{58/(-0.34,-2.39)},
	{59/(0.07,-2.35)},
	{60/(-0.03,-2.07)},
	{61/(0.67,-1.53)},
	{62/(0.48,-1.96)},
	{63/(1.32,-2.30)},
	{64/(1.37,-2.33)},
	{65/(1.66,-1.86)},
	{66/(1.93,-1.55)},
	{67/(1.86,-2.22)},
	{68/(1.51,-1.87)},
	{69/(1.55,-1.29)},
	{70/(2.45,-1.56)},
	{71/(1.90,-1.10)},
	{72/(2.73,-0.95)},
	{73/(2.16,-1.09)},
	{74/(2.81,-0.79)},
	{75/(3.21,-0.87)},
	{76/(3.18,-0.58)},
	{77/(2.51,-0.69)},
	{78/(3.17,-0.21)},
	{79/(2.85,-0.09)},
	{80/(2.79,0.06)}}
\agent{\i}{\x};
\begin{scope}[on background layer]
	\draw (1) -- (3);
	\draw (2) -- (3);
	\draw (1) -- (4);
	\draw (2) -- (4);
	\draw (3) -- (4);
	\draw (2) -- (5);
	\draw (3) -- (5);
	\draw (4) -- (5);
	\draw (3) -- (6);
	\draw (4) -- (6);
	\draw (5) -- (6);
	\draw (4) -- (7);
	\draw (5) -- (7);
	\draw (6) -- (7);
	\draw (5) -- (8);
	\draw (6) -- (8);
	\draw (7) -- (8);
	\draw (6) -- (9);
	\draw (7) -- (9);
	\draw (8) -- (9);
	\draw (7) -- (10);
	\draw (8) -- (10);
	\draw (9) -- (10);
	\draw (8) -- (11);
	\draw (9) -- (11);
	\draw (9) -- (12);
	\draw (10) -- (12);
	\draw (11) -- (12);
	\draw (10) -- (13);
	\draw (11) -- (13);
	\draw (12) -- (13);
	\draw (11) -- (14);
	\draw (12) -- (14);
	\draw (13) -- (14);
	\draw (12) -- (15);
	\draw (13) -- (15);
	\draw (14) -- (15);
	\draw (10) -- (16);
	\draw (13) -- (16);
	\draw (14) -- (16);
	\draw (15) -- (16);
	\draw (14) -- (17);
	\draw (15) -- (17);
	\draw (16) -- (17);
	\draw (15) -- (18);
	\draw (17) -- (18);
	\draw (16) -- (19);
	\draw (17) -- (19);
	\draw (18) -- (19);
	\draw (17) -- (20);
	\draw (18) -- (20);
	\draw (18) -- (21);
	\draw (19) -- (21);
	\draw (20) -- (21);
	\draw (19) -- (22);
	\draw (20) -- (22);
	\draw (21) -- (22);
	\draw (20) -- (23);
	\draw (21) -- (23);
	\draw (21) -- (24);
	\draw (22) -- (24);
	\draw (23) -- (24);
	\draw (23) -- (25);
	\draw (23) -- (26);
	\draw (26) -- (27);
	\draw (25) -- (28);
	\draw (26) -- (28);
	\draw (27) -- (28);
	\draw (26) -- (29);
	\draw (27) -- (29);
	\draw (28) -- (29);
	\draw (27) -- (30);
	\draw (28) -- (30);
	\draw (29) -- (30);
	\draw (1) -- (31);
	\draw (25) -- (31);
	\draw (28) -- (31);
	\draw (29) -- (31);
	\draw (30) -- (31);
	\draw (29) -- (32);
	\draw (30) -- (32);
	\draw (31) -- (32);
	\draw (30) -- (33);
	\draw (31) -- (33);
	\draw (32) -- (33);
	\draw (31) -- (34);
	\draw (32) -- (34);
	\draw (33) -- (34);
	\draw (32) -- (35);
	\draw (33) -- (35);
	\draw (34) -- (35);
	\draw (1) -- (36);
	\draw (33) -- (36);
	\draw (34) -- (36);
	\draw (35) -- (36);
	\draw (34) -- (37);
	\draw (35) -- (37);
	\draw (36) -- (37);
	\draw (35) -- (38);
	\draw (36) -- (38);
	\draw (8) -- (39);
	\draw (36) -- (39);
	\draw (37) -- (39);
	\draw (38) -- (39);
	\draw (38) -- (40);
	\draw (39) -- (40);
	\draw (37) -- (41);
	\draw (38) -- (41);
	\draw (39) -- (41);
	\draw (40) -- (41);
	\draw (40) -- (42);
	\draw (40) -- (43);
	\draw (41) -- (43);
	\draw (42) -- (43);
	\draw (41) -- (44);
	\draw (42) -- (44);
	\draw (43) -- (44);
	\draw (42) -- (45);
	\draw (43) -- (45);
	\draw (44) -- (45);
	\draw (10) -- (46);
	\draw (43) -- (46);
	\draw (44) -- (46);
	\draw (45) -- (46);
	\draw (44) -- (47);
	\draw (45) -- (47);
	\draw (46) -- (47);
	\draw (45) -- (48);
	\draw (46) -- (48);
	\draw (47) -- (48);
	\draw (46) -- (49);
	\draw (47) -- (49);
	\draw (48) -- (49);
	\draw (47) -- (50);
	\draw (48) -- (50);
	\draw (49) -- (50);
	\draw (48) -- (51);
	\draw (49) -- (51);
	\draw (50) -- (51);
	\draw (17) -- (52);
	\draw (49) -- (52);
	\draw (50) -- (52);
	\draw (51) -- (52);
	\draw (19) -- (53);
	\draw (33) -- (53);
	\draw (36) -- (53);
	\draw (47) -- (53);
	\draw (50) -- (53);
	\draw (51) -- (53);
	\draw (41) -- (54);
	\draw (51) -- (54);
	\draw (52) -- (55);
	\draw (53) -- (55);
	\draw (54) -- (55);
	\draw (53) -- (56);
	\draw (54) -- (56);
	\draw (55) -- (56);
	\draw (1) -- (57);
	\draw (39) -- (57);
	\draw (54) -- (57);
	\draw (55) -- (57);
	\draw (56) -- (57);
	\draw (55) -- (58);
	\draw (56) -- (58);
	\draw (57) -- (58);
	\draw (56) -- (59);
	\draw (57) -- (59);
	\draw (58) -- (59);
	\draw (24) -- (60);
	\draw (57) -- (60);
	\draw (58) -- (60);
	\draw (59) -- (60);
	\draw (58) -- (61);
	\draw (59) -- (61);
	\draw (60) -- (61);
	\draw (41) -- (62);
	\draw (59) -- (62);
	\draw (60) -- (62);
	\draw (61) -- (62);
	\draw (60) -- (63);
	\draw (61) -- (63);
	\draw (62) -- (63);
	\draw (22) -- (64);
	\draw (61) -- (64);
	\draw (62) -- (64);
	\draw (63) -- (64);
	\draw (64) -- (64);
	\draw (62) -- (65);
	\draw (63) -- (65);
	\draw (5) -- (66);
	\draw (8) -- (66);
	\draw (63) -- (66);
	\draw (65) -- (66);
	\draw (34) -- (67);
	\draw (54) -- (67);
	\draw (64) -- (67);
	\draw (65) -- (67);
	\draw (66) -- (67);
	\draw (65) -- (68);
	\draw (10) -- (69);
	\draw (53) -- (69);
	\draw (68) -- (69);
	\draw (62) -- (70);
	\draw (68) -- (70);
	\draw (69) -- (70);
	\draw (37) -- (71);
	\draw (68) -- (71);
	\draw (69) -- (71);
	\draw (70) -- (71);
	\draw (48) -- (72);
	\draw (52) -- (72);
	\draw (65) -- (72);
	\draw (69) -- (72);
	\draw (71) -- (72);
	\draw (71) -- (73);
	\draw (72) -- (73);
	\draw (1) -- (74);
	\draw (71) -- (74);
	\draw (72) -- (74);
	\draw (73) -- (74);
	\draw (73) -- (75);
	\draw (53) -- (76);
	\draw (66) -- (76);
	\draw (73) -- (76);
	\draw (74) -- (76);
	\draw (75) -- (76);
	\draw (70) -- (77);
	\draw (74) -- (77);
	\draw (75) -- (77);
	\draw (76) -- (77);
	\draw (20) -- (78);
	\draw (62) -- (78);
	\draw (75) -- (78);
	\draw (2) -- (79);
	\draw (1) -- (80);
	\draw (2) -- (80);
	\draw (3) -- (80);
	\draw (77) -- (80);
	\draw (78) -- (80);
	\draw (79) -- (80);
\end{scope}

\end{tikzpicture}%
	\caption{A sample Erd\H os-R\'enyi network with $N = 80$ nodes, where each node is initialized with $6$ neighbors, and rewired with probability $0.1$.}
	\label{fig:graphNetwork}
\end{figure}
We consider a random Erd\H os-R\'enyi network with $N = 80$ nodes as shown in Figure \ref{fig:graphNetwork}, and apply the algorithms introduced in Section \ref{subsec:gfilterConsensus} with the number of PS iterations $\Gamma = 80$. By suppressing the graph signal component corresponding to the high frequencies, i.e., $\lambda > 0$ for the graph Laplacian $\mathbf{L}$, we achieve the average of the graph signal as the output of the low-pass graph filter. As proposed in Section \ref{subsec:gfilterConsensus}, we use the normalized adjacency matrix $\bar{\mathbf{L}}$ as the shift operator. 
We denote this graph dependent FIR filter as FIR-GDnA, and the graph dependent FIR filter based on the original graph Laplacian as FIR-GDL. As comparisons, two graph independent graph filters, where the frequency responses are designed continuously based on the easily accessed graph properties, i.e., the network size, denoted as FIR-GIDN, and the maximum eigenvalue of the graph Laplacian, denoted as FIR-GIDM, respectively, and the decentralized node-variant graph filter design \cite{a136,a137}, denoted as NV, are considered. The frequency response of the designed FIR graph filters with filter order $K = 12$ are shown in Figure \ref{fig:graphFreRes}. We remark that the FIR-GDnA is a low-pass graph filter, where the low frequency is at $\lambda = 1$. To make sure that the FIR-GDnA achieves the average, a pre-processing and a post-processing step are required, which is illustrated in detail in Appendix \ref{sec:appShiftOperator}.
\begin{figure}[t]
	\begin{minipage}{.45\linewidth}
		\centering
		\input{simulations/graphFreRes}
		\caption{Frequency responses of graph dependent and graph independent FIR graph filters with order $K =12$.}
		\label{fig:graphFreRes}
	\end{minipage}
	\hfill
	\begin{minipage}{.45\linewidth}
		\centering
%
%
\definecolor{mycolor1}{rgb}{0.00000,0.44700,0.74100}%
\definecolor{mycolor2}{rgb}{0.85000,0.32500,0.09800}%
\definecolor{mycolor3}{rgb}{0.92900,0.69400,0.12500}%
\definecolor{mycolor4}{rgb}{0.49400,0.18400,0.55600}%
\definecolor{mycolor5}{rgb}{0.46600,0.67400,0.18800}%
\begin{tikzpicture}

\begin{axis}[%
	width=6.6cm,
	height=4.4cm,
	at={(0cm,0cm)},
	scale only axis,
	xmin=0,
	xmax=14,
	xlabel style={font=\color{white!15!black}},
	xlabel={Filter iteration},
	ymode=log,
	ymin=1e-11,
	ymax=1e+14,
	yminorticks=true,
	ylabel style={font=\color{white!15!black}},
	ylabel={Relative error},
	xmajorgrids,
	ymajorgrids,
	yminorgrids,
	legend columns = 2,
	legend style={at={(0.03,-0.01)}, anchor=south west, legend cell align=left, align=left, draw=none, fill=none}
]
\addplot [color=mycolor1, line width=1pt]
  table[row sep=crcr]{%
0	0.999428850704526\\
1	0.953171420300711\\
2	0.935532200338162\\
3	19.1293115502703\\
4	30.9904613386344\\
5	1228.37945008357\\
6	1485.67462342724\\
7	21823.504885953\\
8	12737.0865181734\\
9	81713.3529884092\\
10	11427.8280328845\\
11	48313.1739703553\\
12	0.000919221683138517\\
};
\addlegendentry{FIR-GDnA(12)}

\addplot [line width = 1pt, mark size = 2pt, color=mycolor2, mark=pentagon*, mark options={fill=mycolor2}, densely dotted]
  table[row sep=crcr]{%
0	79.5219367910916\\
1	81847.739104009\\
2	19668389.5211149\\
3	1542434103.24973\\
4	48551287667.8574\\
5	678886967621.811\\
6	4400937833888.56\\
7	13269342039401.3\\
8	17961808854243.1\\
9	9963175448437.4\\
10	1846679034247.05\\
11	66072585375.5983\\
12	0.00430344251756933\\
};
\addlegendentry{FIR-GDL(12)}

\addplot [line width = 1pt, mark size = 2pt, color=mycolor3, mark=x, mark options={fill=mycolor3}, densely dashed]
  table[row sep=crcr]{%
0	62.2116007854568\\
1	2274.73834361812\\
2	12615.5758831317\\
3	18551.7556991949\\
4	10183.6597255332\\
5	2344.08483280187\\
6	251.034765389742\\
7	12.5571553520084\\
8	1.42827108518858\\
9	1.11184477923422\\
10	1.11394709894772\\
11	1.11375464904365\\
12	1.113758138398\\
};
\addlegendentry{FIR-GIDN(12)}

\addplot [line width = 1pt, mark size = 2pt, color=mycolor4, mark=*, mark options={fill=mycolor4}, densely dashdotted]
  table[row sep=crcr]{%
0	62.2116009393375\\
1	128945.752199936\\
2	45900058.479558\\
3	4568292505.13356\\
4	167040343462.032\\
5	2566768384708.27\\
6	17611940403104.2\\
7	54716342103492.4\\
8	74803429423193.8\\
9	41266929628899.3\\
10	7516694828892.16\\
11	261816401072.147\\
12	0.101159402809651\\
};
\addlegendentry{FIR-GIDM(12)}

\addplot [line width = 1pt, mark size = 2pt, color=mycolor5, mark=triangle*, mark options={fill=mycolor5}, densely dashdotdotted]
  table[row sep=crcr]{%
0	1\\
1	26.2282344952508\\
2	5036.91086120716\\
3	514787.250951484\\
4	96975257.6942602\\
5	522018306.034979\\
6	11352400533.3748\\
7	414992406888.83\\
8	2226960986248.45\\
9	3617037385017.78\\
10	2191459162611.95\\
11	519800668331.202\\
12	37350787446.1691\\
13	724205938.367554\\
14	0.000664783767544569\\
};
\addlegendentry{NV \cite{a136,a137}}

\end{axis}

\end{tikzpicture}%
		\caption{Relative error performance of graph dependent and graph independent FIR graph filters with order $K = 12$.}
		\label{fig:graphError}
	\end{minipage}
\end{figure}

After applying the designed FIR graph filters to the graph signal $\mathbf{x}$, the output $\mathbf{y}$, achieved by equations (\ref{equ:graphFilterxy}) and (\ref{equ:graphFilterH}), is the approximation of the average of the graph signal. Denoting the relative error as
\begin{equation}
	\eta = \frac{\|\mathbf{y} - \bar{\mathbf{x}}\|_2^2}{\|\bar{\mathbf{x}}\|_2^2},
\end{equation}
where $\bar{\mathbf{x}} = \frac{\mathbf{1}^\mathsf{T}\mathbf{x}}{N}$ is the true average of the graph signal, the performance of the designed FIR graph filters are illustrated in Figure \ref{fig:graphError}.
We observe that the graph dependent filter designs, i.e., FIR-GDnA and FIR-GDL, achieve better relative error than the graph independent filter designs, i.e., FIR-GIDN and FIR-GIDM. Furthermore, using the modified normalized graph Laplacian $\bar{\mathbf{L}}$ as the shift operator reduces the total variation during the filtering process as shown in Figure \ref{fig:graphError}, which is important for the stability of the filter in practical implementations with finite numerics. Since the node variant graph filter design enables different filter orders in each node, it requires higher filter orders, thus higher communication overhead, to reach a similar relative error over all nodes.%

\section{Conclusion and Future Work}
\label{sec: conclusion}
In this paper, we propose a decentralized implementation of the online eigendecomposition algorithm for parallel tracking of all eigenvalues of a rank-one modified matrix. Our distributed algorithm is based on parallel averaging consensus protocols and local rational function approximations. Apart from the natural application of our distributed algorithm in the eigenvalue decomposition of the sample covariance matrix and the spectrum computation of the graph Laplacian, we utilize the property of the rank-one modification to perform DoA estimation and DoA tracking, as well as the eigenvalue adaptation of the graph Laplacian in evolving graphs and the graph dependent filter design. All the simulation results show that our decentralized solution of our distributed algorithm converges to the centralized solution at a reduced total communication cost compared to the algorithms facilitated with the distributed power method. Furthermore, the adapted eigenvalues of the graph Laplacian enable the improvement of the total convergence speed in various decentralized algorithms that are based on the decentralized averaging algorithms. 

For future work, applying the rational function approximation approach to decentralized singular value decomposition for non-symmetric matrices is an interesting open problem and worth further investigation. 
\appendices
\section{Deflation Process with the Householder Transformation}\label{sec:appDef}
The rational function approximation approach to perform the eigenvalue decomposition in an online manner is based on the theorem that the eigenvalues of the rank-one update can be found by exploring the roots of a secular function, where we have assumed that the diagonal matrix does not have any repeated eigenvalues as well as that the rank-one update does not contain any zero elements. However, these two assumptions may not hold in practice, especially at the initialization stage where the diagonal matrix has multiple zeros. To overcome this issue, a deflation step is required before the rational function approximation is applied.
\begin{enumerate}[wide]
	\item We first consider the case that the rank-one update $\mathbf{z}$ contains zero element, i.e., $z_k = 0$. In this case, the $k$-th row and column of the diagonal matrix $\boldsymbol{\Lambda}$ are unperturbed by the rank-one matrix $\rho\mathbf{z}\mathbf{z}^\mathsf{H}$, thus the eigenvalue $\bar{\lambda}_k$ of the rank-one update is equal to $\lambda_k$. The remaining eigenvalues $\bar{\lambda}_j$ with $j \neq k$ are the eigenvalues of $\hat{\boldsymbol{\Lambda}} + \rho\hat{\mathbf{z}}\hat{\mathbf{z}}^\mathsf{H}$, where the diagonal matrix $\hat{\boldsymbol{\Lambda}}$ and the column vector $\hat{\mathbf{z}}$ are obtained by removing the $k$-th entry from the diagonal matrix $\boldsymbol{\Lambda}$ and the vector $\mathbf{z}$, respectively.
	
	\item The other case can be reduced to the first case by means of a Householder matrix $\mathbf{H}$, where the Householder matrix $\mathbf{H}$ sets some entries of the vector $\mathbf{z}$ to be zero while the norm of $\mathbf{z}$ remains unchanged, i.e.,
	\begin{equation}
		\mathbf{H}\mathbf{z} = \left\|\mathbf{z}\right\|_2\mathbf{e}_1,
	\end{equation}
	where $\mathbf{e}_1$ is the first column of the identity matrix $\mathbf{I}$.
	
	The Householder matrices in the real domain and complex domain are 
	\begin{equation}
		\mathbf{H}_\mathbb{R} = \mathbf{I} - \frac{2}{\left\|\mathbf{v}\right\|^2}\mathbf{v}\mathbf{v}^\intercal,
	\end{equation}
	and
	\begin{equation}
		\mathbf{H}_\mathbb{C} = \mathbf{I} - \left(1 + \frac{\mathbf{z}^\mathsf{H}\mathbf{v}}{\mathbf{v}^\mathsf{H}\mathbf{z}}\right) \frac{\mathbf{v}\mathbf{v}^\mathsf{H}}{\left\|\mathbf{v}\right\|_2^2},
	\end{equation}		
	respectively \cite{a124}, where $\mathbf{v} = \mathbf{z} - \left\|\mathbf{z}\right\|_2\mathbf{e}_1$.
	The Householder matrix is unitary, i.e., $		\mathbf{H}^\mathsf{H}\mathbf{H} = \mathbf{H}\mathbf{H}^\mathsf{H} = \mathbf{I}$.

	The eigenvalues $\bar{\boldsymbol{\Lambda}}$ and the corresponding eigenvector $\mathbf{V}$ of the rank-one update $(\mathbf{R} = \boldsymbol{\Lambda} + \rho\mathbf{z}\mathbf{z}^\mathsf{H})$ are related as
	\begin{equation}\label{equ:r1}
		\mathbf{V}\bar{\boldsymbol{\Lambda}}\mathbf{V}^\mathsf{H} = \boldsymbol{\Lambda} + \rho\mathbf{z}\mathbf{z}^\mathsf{H}.
	\end{equation}
	Multiplying \eqref{equ:r1} with $\mathbf{H}$ and its Hermitian from the left and the right, respectively, leads to
	\begin{equation}\label{equ:r1H}
		\mathbf{H}\mathbf{V}\bar{\boldsymbol{\Lambda}}\mathbf{V}^\mathsf{H}\mathbf{H}^\mathsf{H} = \mathbf{H}(\boldsymbol{\Lambda} + \rho\mathbf{z}\mathbf{z}^\mathsf{H})\mathbf{H}^\mathsf{H}.
	\end{equation}
	Denote the eigenvalues and the eigenvectors of
	\begin{equation}
		\widetilde{\mathbf{R}} = \mathbf{H}(\boldsymbol{\Lambda} + \rho\mathbf{z}\mathbf{z}^\mathsf{H})\mathbf{H}^\mathsf{H}
	\end{equation}
	as $\widetilde{\boldsymbol{\Lambda}}$ and $\widetilde{\mathbf{V}}$, respectively. Comparing \eqref{equ:r1} and \eqref{equ:r1H}, we conclude that the eigenvalues of $\mathbf{R}$ and $\widetilde{\mathbf{R}}$ coincide, i.e.,
	\begin{equation}
		\widetilde{\boldsymbol{\Lambda}} = \bar{\boldsymbol{\Lambda}},
	\end{equation}
	and their eigenvectors are related as
	\begin{equation}
		\widetilde{\mathbf{V}} = \mathbf{H}\mathbf{V}^\mathsf{H}, \quad \mathbf{V} = \mathbf{H}^\mathsf{H}\widetilde{\mathbf{V}}.
	\end{equation}
	
	Considering a special case where all the eigenvalues are the same, i.e., $\boldsymbol{\Lambda} = \lambda\mathbf{I}$,
	the eigenvalues $\widetilde{\boldsymbol{\Lambda}}$ of $\widetilde{\mathbf{R}}$ are obtained by performing the eigendecomposition of the following rank-one modification with a diagonal matrix
	\begin{equation}
		\widetilde{\mathbf{R}} =\mathbf{H}(\boldsymbol{\Lambda} + \rho\mathbf{z}\mathbf{z}^\mathsf{H})\mathbf{H}^\mathsf{H}
		= \boldsymbol{\Lambda} + \rho\tilde{\mathbf{z}}\tilde{\mathbf{z}}^\mathsf{H},
	\end{equation}
	where $\tilde{\mathbf{z}} = \mathbf{H}\mathbf{z} = \left\|\mathbf{z}\right\|_2\mathbf{e}_1$. A general case where only part of the eigenvalues are the same can be divided into several special cases with a smaller size of the diagonal matrix, the rank-one update vector, and the corresponding Householder matrix.

\end{enumerate}

\section{Normalized Adjacency matrix}\label{sec:appShiftOperator}
We first derive the low frequency and its associated eigenvector of the normalized adjacency matrix $\bar{\mathbf{L}}$ from the graph Laplacian $\mathbf{L}$, whose low frequency is $\lambda = 0$ with the associated eigenvector $\mathbf{1}$, i.e., $\mathbf{L}\mathbf{1} = \mathbf{0}$. Hence,
\begin{equation}
	\begin{aligned}
		\mathbf{D}^{-\frac{1}{2}}\mathbf{L}\mathbf{D}^{-\frac{1}{2}}\mathbf{D}^{\frac{1}{2}}\mathbf{1} &= \mathbf{0}\\
		(\mathbf{I} - \mathbf{D}^{-\frac{1}{2}}\mathbf{L}\mathbf{D}^{-\frac{1}{2}})\mathbf{D}^{\frac{1}{2}}\mathbf{1} &= \mathbf{D}^{\frac{1}{2}}\mathbf{1}\\
		\bar{\mathbf{L}}\mathbf{D}^{\frac{1}{2}}\mathbf{1} &= \mathbf{D}^{\frac{1}{2}}\mathbf{1}.
	\end{aligned}
\end{equation}
We can see that the low frequency is transformed to $\lambda_\mathtt{low} = 1$, and the associated normalized eigenvector is $\mathbf{u}_\mathtt{low} = \mathbf{D}^{\frac{1}{2}}\mathbf{1}/\left\|\mathbf{D}^{\frac{1}{2}}\mathbf{1}\right\|_2$.

We now introduce the pre-processing and post-processing for the FIR-GDnA to achieve the average consensus. Decompose the shift operator as $\bar{\mathbf{L}} = \mathbf{U}\boldsymbol{\Lambda}\mathbf{U}^\mathsf{H}$. Then, from (\ref{equ:graphFilterxy}), (\ref{equ:graphFilterH}) and (\ref{equ:graphFilterResponse}) we have
\begin{equation}
	\begin{aligned}
	\mathbf{y}
	&= 	\sum_{m=0}^{K}h_m\left(\sum_{n=1}^{N}\lambda_n^m\mathbf{u}_n\mathbf{u}_n^\mathsf{H}\right)\mathbf{x}\\
	&= 	\sum_{n=1}^{N}\hat{h}(\lambda_n)\mathbf{u}_n\mathbf{u}_n^\mathsf{H}\mathbf{x}.
	\end{aligned}
\end{equation}
Thus, for the low pass graph filter, the filter output is
\begin{equation}
	\mathbf{y}_\mathtt{low} = \mathbf{u}_\mathtt{low}\mathbf{u}_\mathtt{low}^\mathsf{H}\mathbf{x},
\end{equation}
where the desired output is $	\mathbf{y}_\mathtt{low} \triangleq \frac{1}{N}\mathbf{1}\mathbf{1}^\mathsf{T}\mathbf{x}$. Denote
\begin{equation}
	\mathbf{V} = \frac{\left\|\mathbf{D}^\frac{1}{2}\mathbf{1}\right\|_2}{\sqrt{N}}\mathbf{D}^{-\frac{1}{2}},
\end{equation}
and define the pre-processing and the post-processing as
\begin{equation}
	\tilde{\mathbf{x}} = \mathbf{V}\mathbf{x}, \quad \mathbf{y} = \mathbf{V}^\mathsf{H}\tilde{\mathbf{y}},
\end{equation}
respectively, where $\tilde{\mathbf{y}}=\mathbf{u}_\mathtt{low}\mathbf{u}_\mathtt{low}^\mathsf{H}\tilde{\mathbf{x}}$, we can show that
\begin{equation}
	\begin{aligned}
		\mathbf{y}
		 &=\mathbf{V}^\mathsf{H}\mathbf{u}_\mathtt{low}\mathbf{u}_\mathtt{low}^\mathsf{H}\mathbf{V}\mathbf{x}\\
		 &=\frac{\left\|\mathbf{D}^\frac{1}{2}\mathbf{1}\right\|_2}{\sqrt{N}}\mathbf{D}^{-\frac{1}{2}}\frac{\mathbf{D}^\frac{1}{2}\mathbf{1}}{\left\|\mathbf{D}^\frac{1}{2}\mathbf{1}\right\|_2}\frac{\mathbf{1}^\mathsf{T}\mathbf{D}^\frac{1}{2}}{\left\|\mathbf{D}^\frac{1}{2}\mathbf{1}\right\|_2}\frac{\left\|\mathbf{D}^\frac{1}{2}\mathbf{1}\right\|_2}{\sqrt{N}}\mathbf{D}^{-\frac{1}{2}}\mathbf{x}\\
		 &=\frac{1}{N}\mathbf{1}\mathbf{1}^\mathsf{T}\mathbf{x}.
	\end{aligned}
\end{equation}

Moreover, since the $i$-th node knows the $i$-th entry of $\mathbf{u}_\mathtt{low}$, i.e., $u_{\mathtt{low},i}$ and the the local degree $d_i$, it can compute
\begin{equation}
	\left\|\mathbf{D}^\frac{1}{2}\mathbf{1}\right\|_2 = \frac{\sqrt{d_i}}{|u_{\mathtt{low},i}|}.
\end{equation}
If we assume that the network size $N$ is known, which can also be determined by the PS protocol distributively, and notice that the transformation matrix $\mathbf{V}$ is hermitian and diagonal, we conclude that the pre-processing and the post-processing can be carried out distributively at each node in the network.

\bibliographystyle{IEEEtran}
\bibliography{references.bib}

\end{document}